\documentclass[%
 reprint,
superscriptaddress,
]{revtex4-2}

\usepackage{xcolor}
\usepackage{amsmath}
\usepackage{graphicx}
\usepackage{dcolumn}
\usepackage{multirow}
\usepackage{booktabs}
\usepackage{placeins}

\usepackage{enumitem}

\usepackage[colorlinks=true,linkcolor=blue,citecolor=blue,urlcolor=blue]{hyperref}
\usepackage[nameinlink,capitalise]{cleveref}

\usepackage{subcaption}

\newcommand{\beq}{\begin{equation}}
\newcommand{\eeq}{\end{equation}}
\newcommand{\bea}{\begin{eqnarray}}
\newcommand{\eea}{\end{eqnarray}}

\newcommand{\cb}{{\cal{B}}}
\newcommand{\BB}{{\cal{B}}}

\graphicspath{{pic/}}

\def\Opava{Research Centre for Theoretical Physics and Astrophysics, Institute of Physics, Silesian University in Opava, CZ-74601 Opava, Czech Republic}
\def\Moskva{Faculty of Physics, Moscow State University, 119899, Moscow, Russia}

\begin{document}

\preprint{APS/123-QED}

\title{
Electromagnetic radiation-reaction near black holes: \\ orbital widening and the role of the tail 
}

\author{Bakhtinur Juraev}
 \email{bakhtinur.juraev@gmail.com}\affiliation{\Opava}

\author{Arman Tursunov}
\email{arman.tursunov@physics.slu.cz} \affiliation{\Opava}

\author{Martin Kolo\v{s}}
 \email{martin.kolos@physics.slu.cz}\affiliation{\Opava}

\author{Zden\v{e}k Stuchlík}
 \email{zdenek.stuchlik@physics.slu.cz} \affiliation{\Opava}

\author{Dmitri V. Gal'tsov}\email{galtsov@phys.msu.ru}\affiliation{\Moskva}

\begin{abstract}
We investigate the orbital evolution of a classical charged particle around a Schwarzschild black hole immersed in an external, uniform magnetic field, taking into full account both local radiation‐reaction and the nonlocal “tail” self‐force arising in curved spacetime. Starting from the DeWitt–Brehme equation and its Landau–Lifshitz reduction, we derive analytic expressions for the conservative and dissipative components of the electromagnetic self‐force in both the weak‐field (Newtonian) and strong‐field regimes. By implementing backward‐in‐time integration of the third‐order DeWitt–Brehme equation alongside the second‐order Landau–Lifshitz equation, we demonstrate that the so‐called “orbital widening” effect persists even when the tail term is included, and that for astrophysically realistic charge‐to‐mass ratios the tail contribution to the trajectory is negligible. We further show that this widening is directly controlled by the product of the magnetic field and radiation-reaction parameters and can be captured in the Newtonian limit. Finally, we identify a scaling symmetry showing that simulations with moderate parameter values can accurately represent the dynamics in realistic astrophysical conditions, confirming that orbital widening is a robust phenomenon that can persist even in astrophysical black hole environments.
\end{abstract}

\maketitle

\section{Introduction} \label{sec-intro}

The motion of a charged particle under the influence of its own electromagnetic field, commonly referred to as the radiation reaction or self-force problem, is one of the foundational problems in classical field theory. In curved spacetime, this problem acquires additional complexity due to the nonlocal interaction between the particle and its own field mediated by spacetime curvature. The first rigorous treatment of the electromagnetic self-force in curved backgrounds was provided by DeWitt and Brehme~\cite{DeW-Bre:1960:AP:}, and later refined by Hobbs~\cite{Hob:1968:AP:} to include Ricci curvature contributions. The resulting equation, known as the DeWitt–Brehme (DB) equation, extends the flat-spacetime Lorentz–Dirac (LD) equations by introducing a nonlocal "tail" term that accounts for radiation scattered back onto the particle by the spacetime curvature.

While the DB equation is conceptually complete, its numerical implementation is complicated by both the nonlocal tail term and the presence of third-order derivatives, which can produce unphysical behaviors such as runaway solutions and preacceleration—issues that also arise in the flat-space analogue, the LD equation. To mitigate the latter, a reduced-order method introduced by Landau and Lifshitz~\cite{Landau:1975:CTP2:} replaces the third-order derivatives with a second-order equation by treating radiation reaction as a perturbative correction to the Lorentz force. This approach has proven highly effective for long-time numerical integrations of charged particle motion in strong gravitational and electromagnetic fields. 

Recent interest in this topic has been revitalized by the discovery of the \emph{orbital widening} (OW) effect in magnetized black hole spacetimes~\cite{Tur-Kol-Stu-Gal:2018:APJ:,Tur-Kol-Stu:2018:AN:}. In contrast to the expected inspiral due to radiation losses, radiating charged particles subject to an outward-directed Lorentz force due to external vertical magnetic field can experience a secular increase in orbital radius. This counterintuitive behavior is understood as a result of energy redistribution: although kinetic energy is radiated away, the particle gains potential energy faster due to the outward drift, resulting in a net energy increase at infinity.

The physical robustness of this effect has been questioned in recent work by Santos et al.~\cite{San-Car-Nat:2023:PRD:}. They argued, based on a Newtonian analysis, that the OW effect is an artifact arising from neglecting the electromagnetic tail term in the equations of motion. According to their findings, once the tail term is included, the particle invariably spirals inward, and the OW effect disappears.

In this paper, we revisit the radiation-reaction dynamics of charged particles in the Schwarzschild spacetime immersed in an external, asymptotically uniform magnetic field. We include both local and nonlocal contributions to the electromagnetic self-force and analyze the system using both the full DeWitt–Brehme equations and their reduced-order Landau–Lifshitz counterpart. Our goal is twofold: (i) to investigate the role of the electromagnetic tail in the dynamics of radiating particles complementing our previous studies \cite{Tur-Kol-Stu-Gal:2018:APJ:,Jur-Stuch-Tur-Kol:2024:JCAP:}, and (ii) to assess the physical relevance and persistence of the OW effect under astrophysically realistic conditions.

Our main findings are as follows. First, we confirm that the DeWitt–Brehme and Landau–Lifshitz formulations yield identical trajectories when the tail term is excluded, validating the LL approach for this system. Second, we isolate the tail contribution and confirm its conservative and dissipative character in agreement with analytic results. Third, we derive an energy-balance equation that clarifies the conditions for orbital widening (OW), and demonstrate that the OW effect persists even with the tail included—contrary to prior claims. Fourth, we identify a symmetry that maps trajectories across many orders of magnitude in parameters of the model, establishing the astrophysical relevance of our findings. Finally, we quantify the magnitudes of the relevant forces in realistic settings involving electrons near stellar-mass black holes. We also demonstrate, via numerical simulations, that the OW effect occurs even at astrophysically relevant values of the parameters, confirming its relevance for magnetized compact-object environments.

\subsection{Parameter setup} \label{sec-intro-setup}

In this subsection, we summarize the assumptions, coordinate system, and parameter definitions that will be used throughout the paper. Our setup follows earlier works on charged particle motion in magnetized black hole spacetimes~\cite{Kol-Stu-Tur:2015:CLAQG:,Stu-Kol:2016:EPJ:,Tur-Stu-Kol:2016:PRD:,Tur-Kol-Stu-Gal:2018:APJ:,Stuch-Kol-Tur:2024:Universe:,2024JHEAp..44..500S,2020Univ....6...26S}. We will use solutions derived in \citep{Smi-Wil:1980:PRD:,Gal:1982:JPMG:,DeWitt-DeWitt:1964:PPF:} for the tail term. 

We consider a Schwarzschild black hole of mass $M$, described in standard spherical coordinates $(t, r, \theta, \phi)$ by the line element
\begin{equation}
    ds^{2} = - f(r) \, dt^{2} + \frac{dr^{2}}{f(r)} + r^{2} \, d\theta^{2} + r^{2} \sin^{2} \theta \, d\phi^{2},
\label{Schwarschild_metric}
\end{equation}
where the lapse function is 
\begin{equation}
    f(r) = 1 - \frac{2M}{r}.
\end{equation} 
The black hole is assumed to be immersed in an external, asymptotically uniform magnetic field, described by the well-known Wald solution~\cite{Wald:1974:PHYSR4:}. In this configuration, the only nonvanishing component of the electromagnetic four-potential is
\begin{equation} \label{vecpot}
    A_{\phi} = \frac{B}{2} r^{2} \sin^{2} \theta,
\end{equation}
where $B$ is the magnetic field strength as measured at spatial infinity.

The conserved energy and generalized azimuthal angular momentum per unit mass of a charged particle moving in this background are associated with the timelike and azimuthal Killing vectors, and are given by
\begin{eqnarray}
    \mathcal{E} &\equiv& \frac{E}{m} = f(r) \, \frac{dt}{d\tau}, \label{Energy} \\
    \mathcal{L} &\equiv& \frac{\pi_{\phi}}{m} = r^{2} \sin^{2} \theta \left( \frac{d\phi}{d\tau} + \frac{qB}{2m} \right), \label{Ang-mom}
\end{eqnarray}
where $q$ and $m$ are the particle's charge and mass, respectively, and $\tau$ is the proper time.

Following~\cite{Tur-Kol-Stu-Gal:2018:APJ:}, we introduce two key dimensionless parameters:
\begin{align}
    k &\equiv \frac{2 q^{2}}{3m}, \quad \text{radiation–reaction parameter,} \label{k-param} \\
    \mathcal{B} &\equiv \frac{q B}{2m}, \quad \text{magnetic coupling parameter.} \label{B-param}
\end{align}
The parameter $k$ is always positive, while $\mathcal{B}$ can be both positive and negative, depending on the orientation of the magnetic field with respect to the direction of the motion. 
A useful symmetry of the system is the interchange $(\mathcal{L},\mathcal{B}) \leftrightarrow (-\mathcal{L}, -\mathcal{B})$. Without loss of generality, we take positive angular momentum, ${\cal L}>0$, to correspond to counterclockwise motion, which is assumed in all numerical plots throughout this paper. 
 This allows us to classify the dynamics into two qualitatively distinct regimes:
\begin{itemize}
    \item $\mathcal{B} > 0$: the Lorentz force is repulsive, pushing the particle away from the black hole; 
    \item $\mathcal{B} < 0$: the Lorentz force is attractive, pulling the particle toward the black hole.
\end{itemize}
Throughout this paper, we will restrict to equatorial motion ($\theta = \pi/2$), where the dynamics simplify due to symmetry, and where the OW effect is most easily analyzed. 
We use the spacetime signature $(-,+,+,+)$ and the $G = c = 4 \pi \epsilon_{0} = 1$ system of geometric units throughout the paper. We use the constants in the Gaussian-CGS system of units explicitly for expressions with astrophysical relevance. In the graphics, we use red color to represent cases with a tail term, while black color is to denote without a tail term. The range of Greek indices is from 0 to 3.

The remainder of the paper is organized as follows. In Sec.~\ref{Rad_particle_curved_spacetime}, we review the DeWitt–Brehme and Landau–Lifshitz equations in Schwarzschild spacetime with a uniform magnetic field, and give explicit expressions for the self-force components, including the tail term. In Sec.~\ref{sec:numresults}, we compare the two formalisms numerically and study the pure tail contribution. Section~\ref{sec:OWeffect} focuses on the orbital widening effect, both in the Newtonian limit and in full general relativity. In Sec.~\ref{sec:astro}, we discuss the scaling symmetry and force hierarchy, and we assess the astrophysical relevance of the OW effect. We conclude in Sec.~\ref{Conclusions}.

\section{Radiation reaction in curved spacetime} \label{Rad_particle_curved_spacetime}

\subsection{DeWitt-Brehme equation}

More than half a century ago, DeWitt and Brehme derived the first rigorous expression for the electromagnetic self‑force in curved backgrounds \cite{DeW-Bre:1960:AP:}. Hobbs subsequently completed their analysis \cite{Hob:1968:AP:} by clarifying the contribution of the Ricci curvature to the equations. The motion of a point charge in a curved spacetime is thus governed by 
\bea
\begin{split}
 \frac{D u^{\mu}}{d \tau} = \frac{q}{m} F^{\mu}_{\  \nu} u^{\nu} + \frac{2 q^{2}}{3m} \left( \frac{D^{2} u^{\mu}}{d \tau^{2}} + u^{\mu} u_{\nu} \frac{D^{2}u^{\nu}}{d \tau^{2}}\right) +  \\  \frac{q^{2}}{3m} (R^{\mu}_{\ \lambda}u^{\lambda} + R^{\nu}_{\ \lambda} u_{\nu} u^{\lambda} u^{\mu} ) + \mathcal{F}^{\mu}_{tail}\,\,, 
\end{split}
\label{eqmoDWBH} \label{LD}
\eea
where $q, m, u^\alpha$, are the charge, mass, and the four-velocity of a test particle, $F_{\mu\nu}$ is the electromagnetic Faraday tensor, $D$ denotes a covariant derivative. The first row represents the classical, well-known Lorentz–Dirac equation; in flat spacetime, the covariant derivatives reduce to ordinary derivatives. The second row contains the curvature-induced corrections together with the nonlocal tail term, yielding the complete DeWitt–Brehme equation in curved spacetime. The term containing the Ricci tensor vanishes in the vacuum metrics, so it is irrelevant in our case.  

The last term of Eq.~(\ref{eqmoDWBH}) is the so-called "tail" term given by derivatives of Green's function 
\begin{equation}
  \mathcal{F}^{\mu}_{tail} = \frac{2 q^{2}}{m} u_{\nu} \int_{-\infty}^{\tau-0^+}     
D^{[\mu} G^{\nu]}_{ + \lambda'} \bigl(z(\tau),z(\tau')\bigr)   
u^{\lambda'} \, d\tau'.
\label{tail}
\end{equation}
The integral in this term is evaluated over the entire past history of the charged particle with primes indicating its prior positions. The existence of the "tail" integral implies that the radiation reaction in curved spacetime has non-local nature. A detailed derivation of the equations of motion for a radiating charged particle in curved spacetime is given, for example, in~\cite{Poisson:2004:LRR:}. Its flat‑space analogue can be found in~\cite{Spohn:2000:EPL:}.

From the normalization condition of the four-velocity, 
\begin{equation}
    u_{\alpha} u^{\alpha} = -1,
\end{equation}
one obtains the following relations
\begin{equation}
    u_{\alpha} \dot{u}^{\alpha} = 0, 
    \qquad 
    u_{\alpha} \ddot{u}^{\alpha} = - \dot{u}_{\alpha} \dot{u}^{\alpha}.
    \label{con_four_vel}
\end{equation}
%
Here dot indicates the covariant derivative. Thus, in vacuum metrics, one can rewrite Eq.~(\ref{LD}) using (\ref{con_four_vel}) in the following form
\begin{eqnarray}
   && \frac{D u^{\mu}}{d \tau} = \frac{q}{m} F^{\mu}_{\ \nu} u^{\nu} + \nonumber \\ 
  && \frac{2 q^{2}}{3m} \left( \frac{D^{2} u^{\mu}}{d \tau^{2}} - \right. 
   \left. \frac{D u^{\nu}}{d \tau} \frac{D u_{\nu}}{d \tau} u^{\mu} \right) 
    +   \mathcal{F}^{\mu}_{tail} \, . 
    \label{LD-2}
\end{eqnarray}
In contrast to the original form, this rewritten version of the DeWitt–Brehme equation now contains only a single third-order derivative term, which is expected to reduce numerical errors during integration. 

The term $D^{2} u^{\mu} / d \tau^{2}$ can be rewritten in the following form \cite{Tur-Kol-Stu-Gal:2018:APJ:}
\begin{align}
   \frac{D^{2} u^{\mu}}{d \tau^{2}} &= \frac{d^{2} u^{\mu}}{d \tau^{2}} + \left( \frac{\partial \Gamma^{\mu}_{\alpha \beta}}{\partial x^{\gamma}} u^{\gamma} u^{\beta} + 3 \Gamma^{\mu}_{\alpha \beta} \frac{d u^{\beta}}{d\tau} + \right. \nonumber \\
   &\quad + \left. \Gamma^{\mu}_{\alpha \beta} \Gamma^{\beta}_{\rho \sigma} u^{\rho} u^{\sigma} \right) u^{\alpha}, 
\end{align}
where $\Gamma^{\mu}_{\alpha \beta}$ are Christoffel symbols. 

\subsubsection*{Explicit form of the DeWitt-Brehme equations}
We now present the explicit components of the DeWitt–Brehme equation for a particle in Schwarzschild spacetime immersed in an asymptotically uniform magnetic field of the strengths $B$, given by the Wald's solution to electromagnetic fields~\cite{Wald:1974:PHYSR4:}, and defined in (\ref{vecpot}). 
 
We fix the plane of motion to the equatorial plane, $\theta = \pi/2$, and, without loss of generality, set the black hole mass to unity, $M = 1$. Thus, the $\theta$-equation vanishes, and the non-vanishing components of the equations of motion, Eq.~(\ref{LD-2}) take the form 
\begin{widetext}

\begin{equation}
\begin{split}
   \frac{d u^t}{d \tau } &= - \frac{2 u^{r} u^{t}}{f r^{2}}  
   - \frac{k}{f^{3} r^{4}} \bigg\{  
   f^{3} r^{3} (u^{\phi})^{2} u^{t} - f (u^{r})^{2} u^{t}  
   + 4 f (1 + f r) u^{t} (u^{r})^{2}  \\ 
   &\quad - f^{3} (u^{t})^{3}  
   - 3 f^{2} r^{2} (a^{t} u^{r} + a^{r} u^{t})   
    + u^{r} \bigg[  
   f^{3} r^{4} (a^{\phi} r + 2 u^{\phi} u^{r})^{2}  
   - f^{2} (a^{t} f r^{2} + 2 u^{r} u^{t})^{2}  \\ 
   &\quad - (a^{r} f r^{2} - f^{2} r^{3} (u^{\phi})^{2}  
   - (u^{r})^{2} + f^{2} (u^{t})^{2} )^{2} \bigg]  
   - f^{3} r^{4} \frac{d a^{t}}{d \tau} \bigg\}  
   + \mathcal{F}^{t}_{tail},
\label{EOM1LD}
\end{split}
\end{equation}

\begin{equation}
\begin{split}
   \frac{d u^r}{d \tau } &= f r u^{\phi}  \{2\mathcal{B} + u^{\phi}  \} - \frac{1}{r^{2}} - (u^{\phi})^{2}  - k \bigg\{ 3fr a^{\phi} u^{\phi} + \frac{3  a^{r} u^{r}}{f r^{2}} 
   + 2  f  u^{r}(u^{\phi})^{2} - \frac{u^{r}  (u^{\phi})^{2}}{r}  \\  &\quad + 
   \frac{(2 + fr) u^{r}  (u^{\phi})^{2}}{r} 
   -   \frac{(u^{r})^{3}}{f^{2}  r^{4}}  -  \frac{2 (1 + 
      fr)(u^{r})^{3}}{f^{2}  r^{4}}  - \frac{3  f  a^{t}  u^{t}}{r^{2}} 
      - \frac{u^{r}  (u^{t})^{2}}{r^{4}}  
      + \frac {2 (fr-1) u^{r}(u^{t})^{2}}{r^{4}} \\ 
   &\quad + u^{r} \bigg[  (a^{\phi} + 2  u^{\phi}  u^{r})^{2} 
      -f\left(a^{t} + \frac {2u^{r} u^{t}}{fr^{2}} \right)^{2}  
      + \frac{((u^{r})^{2} + f  (f  r^{3}  (u^{\phi})^{2} - 
        f  (u^{t})^{2}))^{2}}{f^{3}  r^{4}} \bigg]  
      - \frac{d a^{r}}{d\tau} \bigg\}  +  \mathcal{F}^{r}_{tail},
\label{EOM2LD}
\end{split}
\end{equation}

\begin{equation}
\begin{split}
  \frac{d u^{\phi }}{d\tau } &= - \frac{2 u^{r} (\mathcal{B} + u^{\phi})}{r}  
  - \frac{k}{f^{3} r^{4}} \bigg\{  
  f^{4} r^{4} (u^{\phi})^{3}  
  + f^{2} r u^{\phi} (u^{r})^{2}  
  - 3 f^{3} r^{3} (a^{r} u^{\phi} + a^{\phi} u^{r}) - f^{4} r u^{\phi} (u^{t})^{2}  \\ 
  &\quad   
  + u^{\phi} \bigg[  
  f^{3} r^{4} (a^{\phi} r + 2 u^{\phi} u^{r})^{2}  
  - f^{2} (a^{t} f r^{2} + 2 u^{r} u^{t})^{2}  + (a^{r} f r^{2} - f^{2} r^{3} (u^{\phi})^{2}  
  - (u^{r})^{2} + f^{2} (u^{t})^{2})^{2} \bigg]  
  - f^{3} r^{4} \frac{d a^{\phi}}{d\tau} \bigg\}  
  + \mathcal{F}^{\phi}_{tail},
\label{EOM3LD}
\end{split}
\end{equation}

\end{widetext}
where $a^{\alpha} = d u^{\alpha}/ d \tau$ is coordinate (non-covariant) four-acceleration. 

The explicit components of the electromagnetic self-force $\mathcal{F}^{\mu}_{tail}$ will be specified below in Eqs.~(\ref{tail_r}), (\ref{tail_Gal_t})-(\ref{tail_Gal_phi}).

\subsection{Covariant Laundau-Lifshitz reduction} 

The DeWitt–Brehme equation, as a covariant generalization of the Lorentz–Dirac equation to curved spacetime, poses several challenges for numerical integration, particularly due to the local third-order derivative terms. These include runaway solutions -- unphysical trajectories where the particle’s acceleration grows exponentially even in the absence of external forces, and the pre-acceleration, where the particle begins to respond to a force before it is applied, violating causality. These issues arise from the third-order nature of the equation, which introduces nonphysical degrees of freedom. 

The runaway solutions can be avoided by integrating the equations backward in time, as proposed in \cite{Bay-Hus:1976:PRD:}. This approach ensures that the solution asymptotically approaches the physical branch.  
Later we test it numerically for Schwarzschild spacetime case.

However, the issue of pre-acceleration still persists, even when using the backward integration method. To resolve this, one can adopt a covariant version of the reduced-order formulation based on the Landau–Lifshitz approach \citep{Landau:1975:CTP2:}. This method treats the radiation reaction as a perturbative correction to the Lorentz force, effectively replacing the third-order equation with a causal second-order one that avoids both runaway and pre-accelerated solutions. It thus offers a more stable and physically consistent framework for numerical implementation. The reduced-order equations of motion take the form (see, \cite{Tur-Kol-Stu-Gal:2018:APJ:})
\begin{equation}
\begin{split}
 \frac{D u^{\mu}}{d \tau}
  =  \frac{q}{m} F^{\mu}_{\,\,\,\nu} u^{\nu} + k \frac{q}{m} F^{\mu}_{\,\,\,\nu ; \alpha} u^\nu u^\alpha +
\\
 k \frac{q^{2}}{m^{2}} F^{\mu}_{\,\,\,\nu}
F^{\nu}_{\,\,\,\alpha} u^{\alpha} 
+ k\frac{q^{2}}{m^{2}} F_{\alpha\beta} F^{\beta}_{\,\,\,\sigma} u^\sigma u^\mu u^{\alpha} + 
  \mathcal{F}^{\mu}_{tail} \, . 
\label{LL}
\end{split}
\end{equation} 
where $k$ is a radiation-reaction parameter defined by (\ref{k-param}) and semicolon denotes the covariant derivative. 
A similar equation was obtained by Quinn and Wald (see, eq.~(26) in \cite{Quinn-Wald:1997:PRD:}), using Hadamard expansions.

In the equation above, the first term on the right-hand side corresponds to the Lorentz force, whose influence on charged-particle dynamics in curved spacetime has been extensively studied in literature (see, e.g., \cite{Kol-Stu-Tur:2015:CLAQG:,Tur-Stu-Kol:2016:PRD:}). The next three terms represent the local part of the radiation-reaction force. Their effects on particle motion have been analyzed in both Schwarzschild and Kerr spacetimes, in various field settings~\cite{Jur-Stuch-Tur-Kol:2024:JCAP:,Tur-Kol-Stu-Gal:2018:APJ:,Kol-Tur-Stu:2021:PRD:}. The final term is, as usually, the nonlocal tail term, which corresponds to the particle’s self-force.  

\subsubsection*{Explicit form of the Landau-Lifshitz equations}
Similar to the previous subsection, we now write the components of the equations of motion in explicit form (see, \cite{Tur-Kol-Stu-Gal:2018:APJ:}). Fixing the motion to the equatorial plane, $\theta = \pi/2$, and using the four-potential given in (\ref{vecpot}), together with the parameterizations introduced in (\ref{k-param}) and (\ref{B-param}), we obtain from (\ref{LL}) the following equations 
\begin{widetext}
\bea
\frac{d u^t}{d \tau } &=& - \frac{2 u^{r} u^{t}}{f r^{2}} - \frac{2 k \mathcal{B} u^{t}}{r} \{ 2 \mathcal{B} r f[f (u^{t})^{2} - 1] - u^{\phi} \} + \mathcal{F}^{t}_{tail},
\label{EOM1LL} \\ 
\frac{d u^r}{d \tau } &=& - \frac{1}{r^{2}} + 2 \BB f r u^{\phi} + (u^{\phi})^{2}[fr - 1] - \frac{2 k \mathcal{B} u^{r}}{r} \{ 2 \BB r f^{2} (u^{t})^{2} - u^{\phi} \} + \mathcal{F}^{r}_{tail},
\label{EOM2LL} \\
\frac{d u^{\phi }}{d\tau } &=& - \frac{2 u^{r} ( u^{\phi} + \BB)}{r} + \frac{2 k \mathcal{B}}{r^{3}} \{ r^{2} (u^{\phi})^{2} - 2 \mathcal{B} r^{3} f^{2} (u^{t})^{2} u^{\phi} + 1 \} + \mathcal{F}^{\phi}_{tail}.
\label{EOM3LL} 
\eea
\end{widetext}

One can see that this set of equations is significantly simpler than the higher-order DeWitt-Brehme equations (\ref{EOM1LD})–(\ref{EOM3LD}). We will further demonstrate that numerically integrating the system of equations (\ref{EOM1LD})–(\ref{EOM3LD}) yields exactly the same trajectories as those obtained from integrating the reduced order equations of motion (\ref{EOM1LL})–(\ref{EOM3LL}).

\subsection{Electromagnetic self-force}
Solving the electromagnetic self-force, given by the tail term in Eq.~(\ref{tail}), is a significant technical challenge. Although the self-force in curved spacetime is fundamentally non-local, as it depends on an integral over the entire past history of the particle, there exist special cases where this non-local expression can be rewritten in a compact local form. Such local expressions arise under certain approximations for highly symmetric spacetimes and particle motions, as in the case of circular orbits in Schwarzschild and Kerr spacetimes \cite{Smi-Wil:1980:PRD:,DeWitt-DeWitt:1964:PPF:,Gal:1982:JPMG:}. In these scenarios, the tail integral can be explicitly evaluated or expanded as a power series, leading to analytic, local formulas for the self-force. %

It is important to note that the tail term, which is independent from external fields, consists of conservative and dissipative components, which arise from the formal decomposition of the self-force using time-symmetric and time-antisymmetric Green's functions \cite{Det-Whi:2004:LRR:,Poisson:2004:LRR:}. This method was first presented by Wald in the context of electromagnetic self-force~\cite{Wald:1978:PRD:}.

A solution to Eq.~(\ref{tail}) in the Newtonian approximation was originally derived by DeWitt and DeWitt~\cite{DeWitt-DeWitt:1964:PPF:}, and was recently employed in~\cite{San-Car-Nat:2023:PRD:} to study the motion of a charged particle around a massive object within the Newtonian limit.

We now introduce the explicit components of the self-force in Schwarzschild spacetime, which will be further used in our numerical analysis of charged particle motion.

\subsubsection{Conservative part}
The conservative component of the self-force in Schwarzschild spacetime was derived by Smith and Will~\cite{Smi-Wil:1980:PRD:}, and takes the explicit radial form 
\begin{equation}
    \mathcal{F}^{r}_{tail} = \frac{3 k}{2 r^{3}} \left( 1 - \frac{2M}{r} \right)^{1/2}.
\label{tail_r}
\end{equation}
It is important to note that the conservative part of the self-force acts as a repulsive force, pushing the particle away from the black hole \cite{Poi-Pou-Veg:2011:LRR:}. Such behavior might appear counter-intuitive, given that the overall effect of radiation reaction is expected to remove energy and angular momentum from the particle’s orbit, causing orbital decay. Haas \cite{Haas:2008:PHDThesis:} highlighted this peculiar feature of the conservative self-force in the scalar field case. 

This interplay between energy dissipation and radial directionality underscores the complexity of self-interaction effects in curved spacetime. Our numerical results presented in later sections confirm and further quantify these effects for charged particles undergoing electromagnetic self-force in Schwarzschild spacetime.

\subsubsection{Dissipative part}

The dissipative components of the electromagnetic self-force are the $t$ and $\phi$ components of $\mathcal{F}_{\mu}^{tail}$.  
 These components for Schwarzschild spacetime can be derived as a special limit ($a=0$) of the more general Kerr solution obtained by Gal'tsov \cite{Gal:1982:JPMG:} 
\begin{eqnarray}
    \mathcal{F}^{tail}_{t} &=& \frac{k \Omega^{2}_{\phi}}{r^{4}} (r^{6} \Omega^{2}_{\phi} + 4) , \label{tail_Gal_t} \\
      \mathcal{F}^{tail}_{\phi} &=&  - \frac{k \Omega_{\phi}}{r^{4}} (r^{6} \Omega^{2}_{\phi} + 4) ,  \label{tail_Gal_phi}
\end{eqnarray}
where $\Omega_{\phi} = u^{\phi}/u^{t}$ is an angular frequency of the particle. 
One can see that the dissipative components are related through a simple algebraic relation (see for more details \citep{War-Bara:2010:PRD:,Gal:1982:JPMG:})
\begin{equation}
\mathcal{F}_{t}^{tail} + \mathcal{F}_{\phi}^{tail}  \Omega_{\phi} = 0.
\end{equation}
The analytical results obtained by Gal'tsov have been thoroughly tested and confirmed numerically by Warburton and Barack \cite{War-Bara:2010:PRD:}, demonstrating excellent agreement between analytic and numeric computations.

\section{Numerical results} \label{sec:numresults}

\FloatBarrier
\begin{figure*}[htbp]
    \centering
    \fbox{%
        \begin{minipage}{0.98\textwidth}
            \centering
    \includegraphics[width=0.95\hsize]{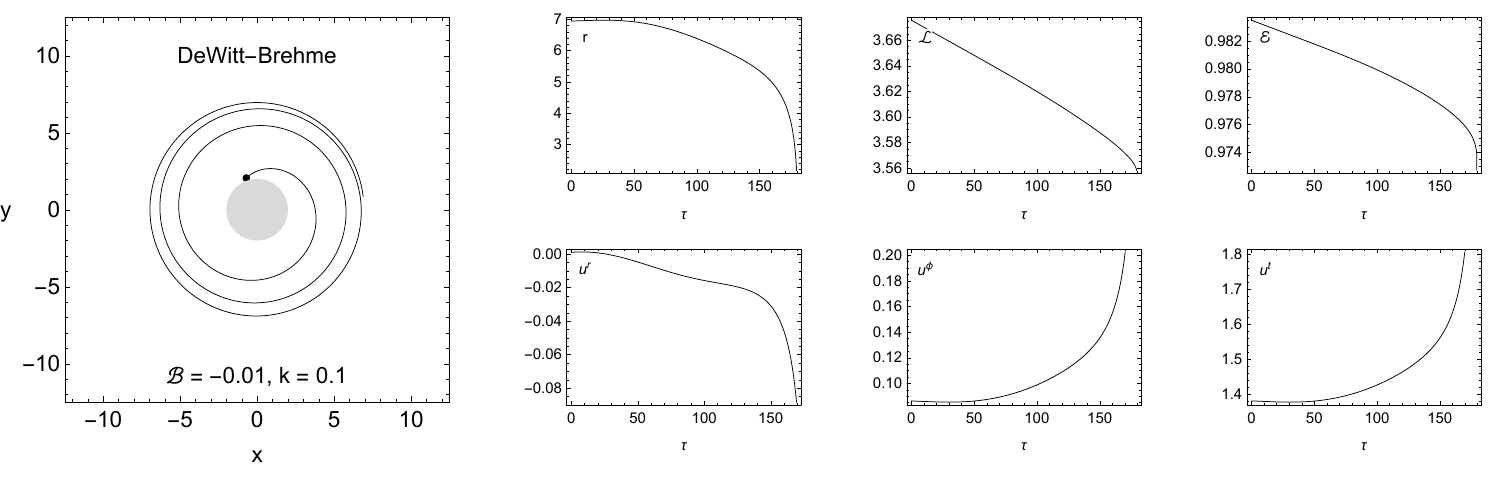}
    \includegraphics[width=0.95\hsize]{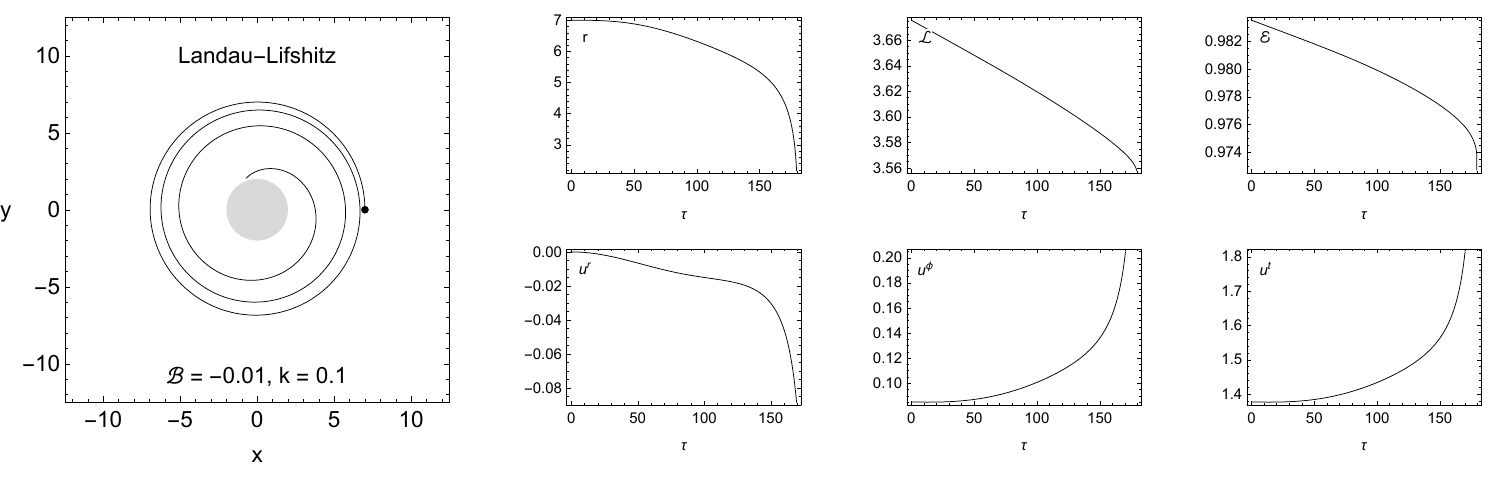}
        \end{minipage}%
    }
    \vspace{2mm}
    \fbox{%
        \begin{minipage}{0.98\textwidth}
            \centering
    \includegraphics[width=0.95\hsize]{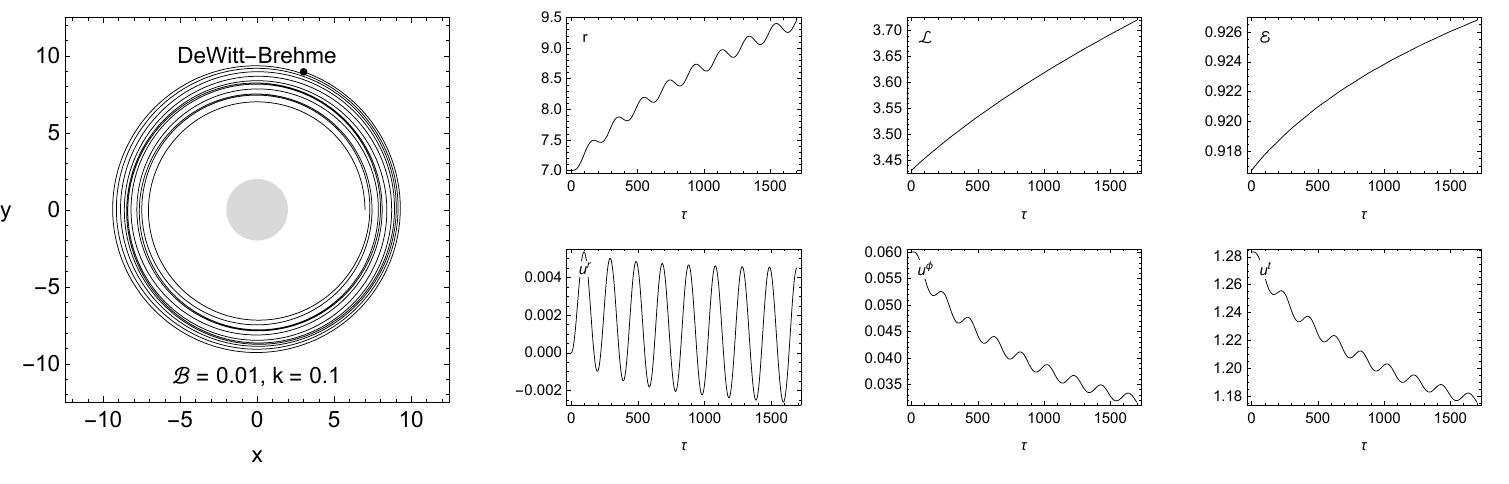}
    \includegraphics[width=0.95\hsize]{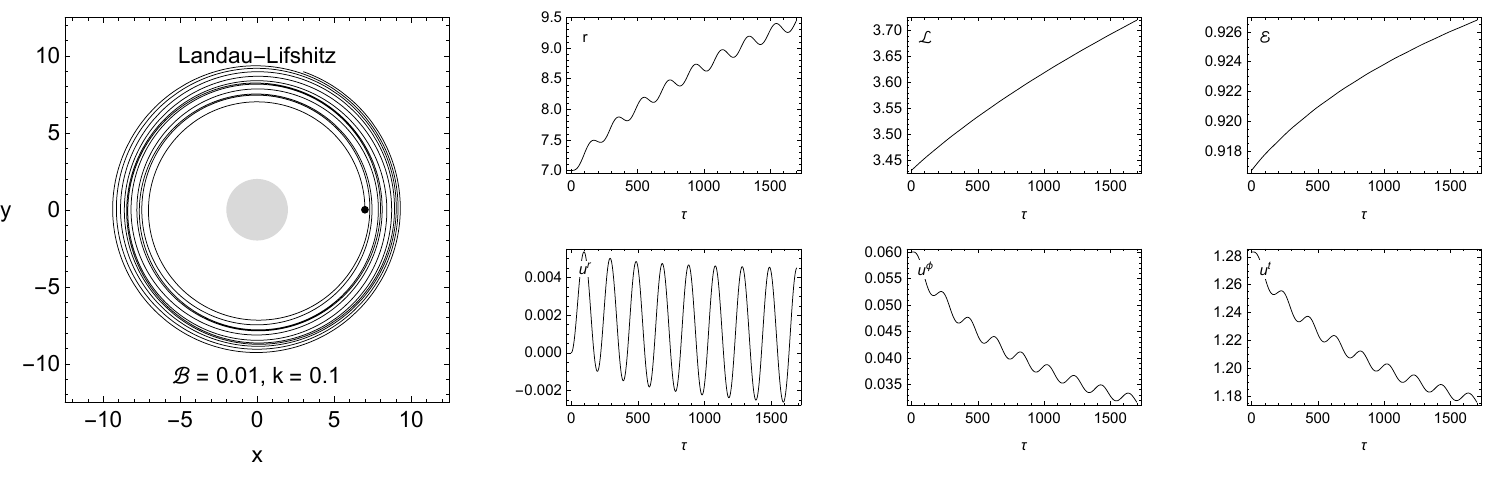}
        \end{minipage}%
    }
    \caption{Comparison of trajectories obtained by integrating the DeWitt-Brehme (DB) Eqs.  (\ref{EOM1LD})--(\ref{EOM3LD}) and the Landau-Lifshitz (LL) Eqs. (\ref{EOM1LD})--(\ref{EOM3LD}) for both negative $\mathcal{B}$ (top two rows) and  positive $\mathcal{B}$ (bottom two rows). The tail term is neglected in all cases. 
    DB is integrated backwards in time, so that the  starting point of the LL trajectory is the ending point of the DB trajectory in our numerical settings. Particles are spiraling out in case of $\mathcal{B}>0$. }
    \label{Fig7}
\end{figure*}

\begin{figure}
\centering
\includegraphics[width=1\hsize]{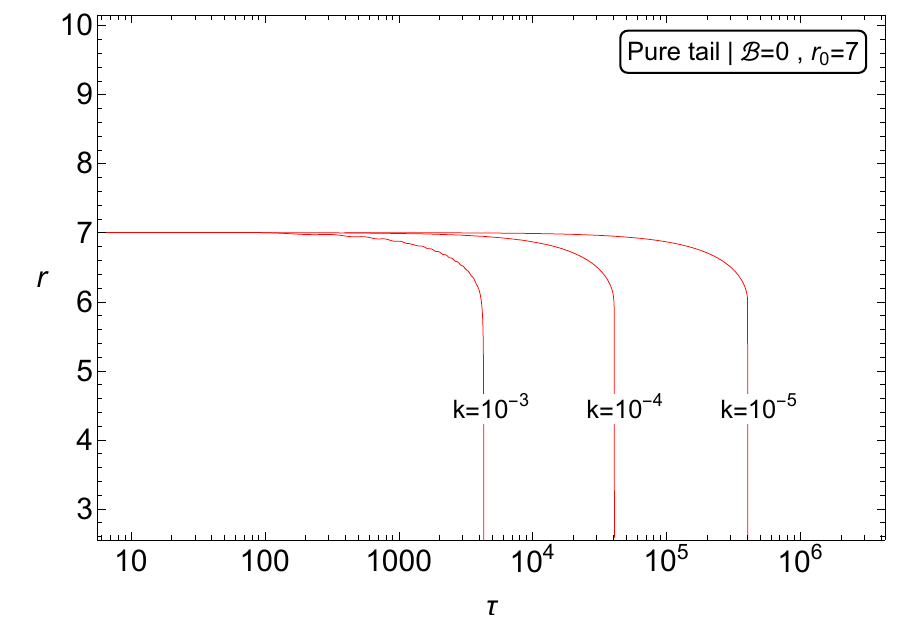}
\caption{Influence of the tail term on the charged particle motion for different values of the radiation parameter $k$. The starting point of the particle is $r_{0}=7$. Lorentz and radiation-reaction forces are set to zero.}
\label{Fig18}
\end{figure}

\begin{figure*}[htbp]
\centering
\includegraphics[width=1\hsize]{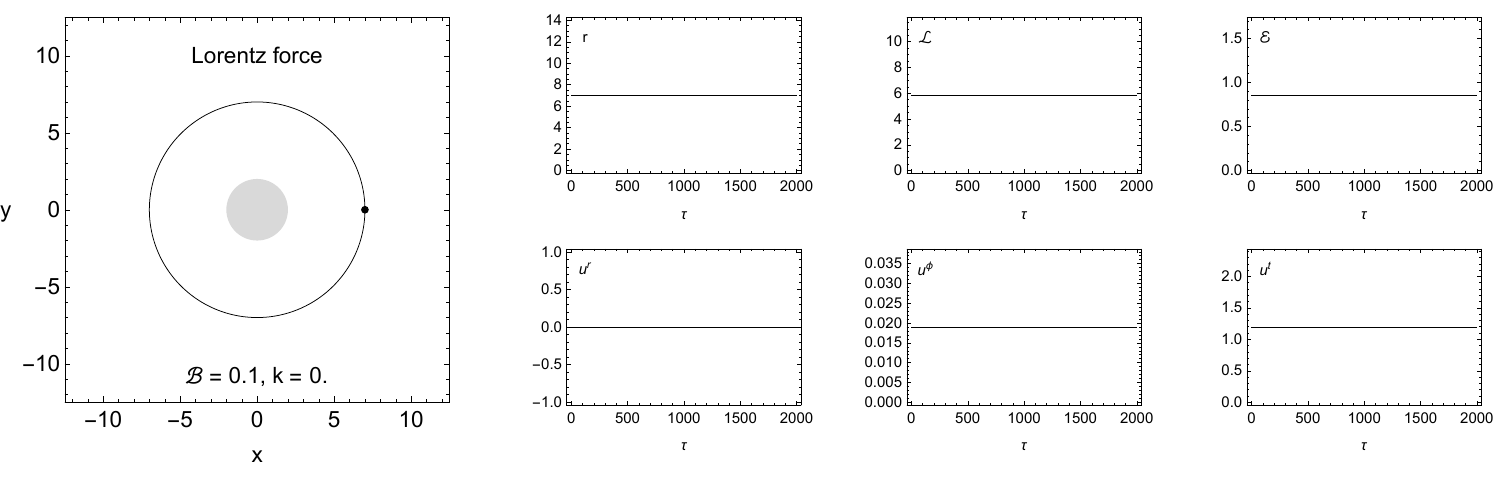}
\includegraphics[width=1\linewidth]{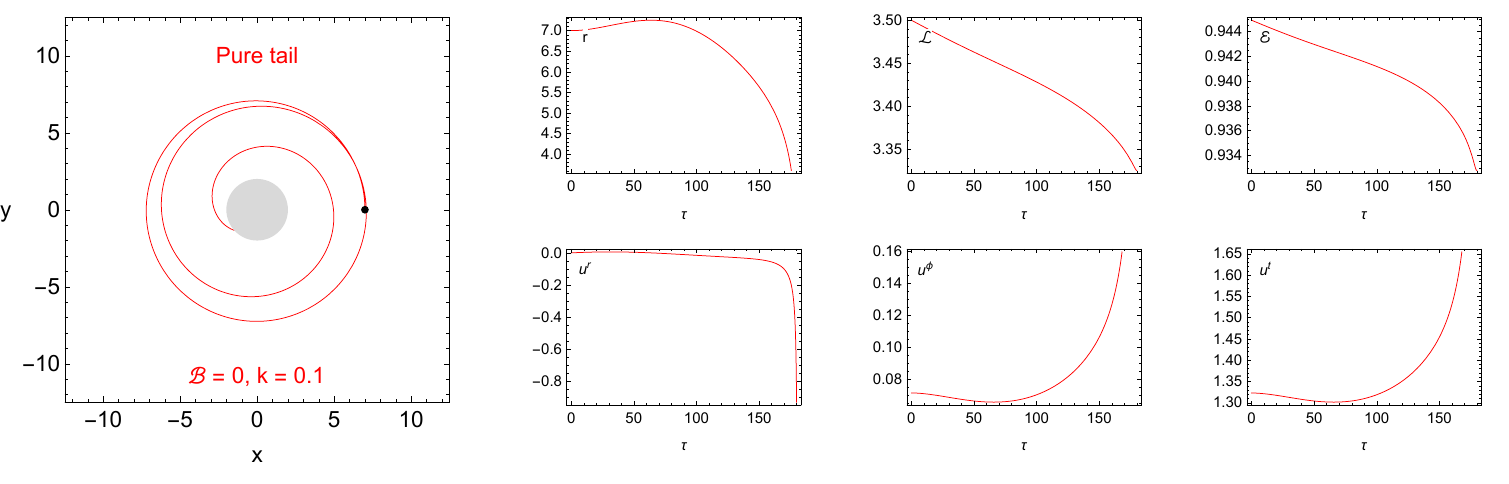}
\includegraphics[width=1\linewidth]{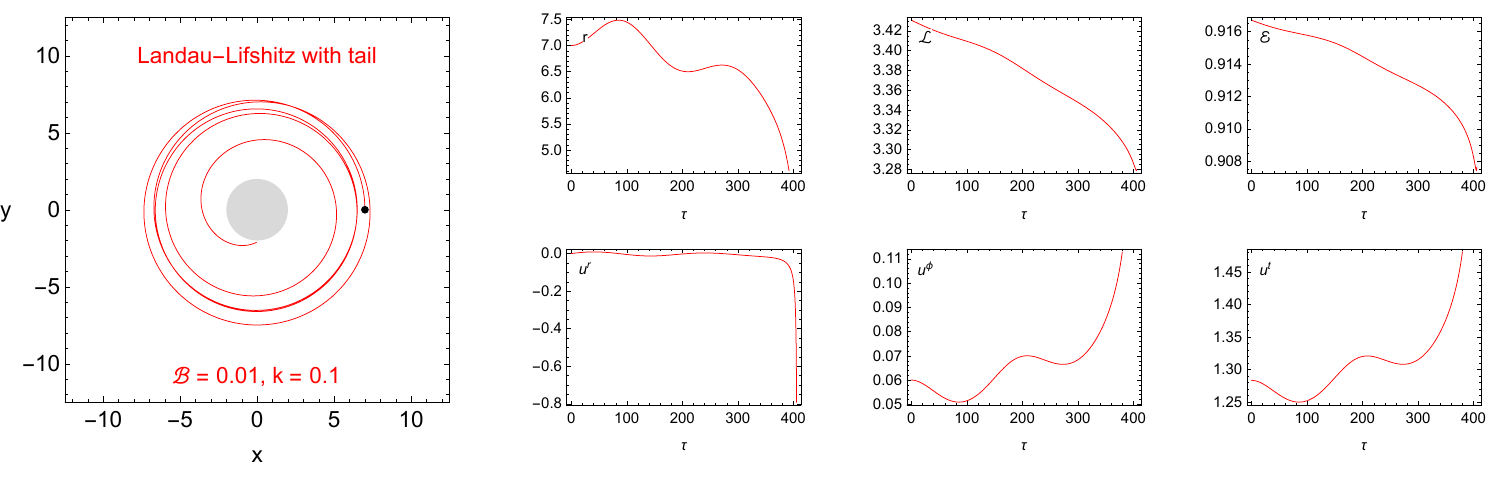}
\includegraphics[width=1\linewidth]{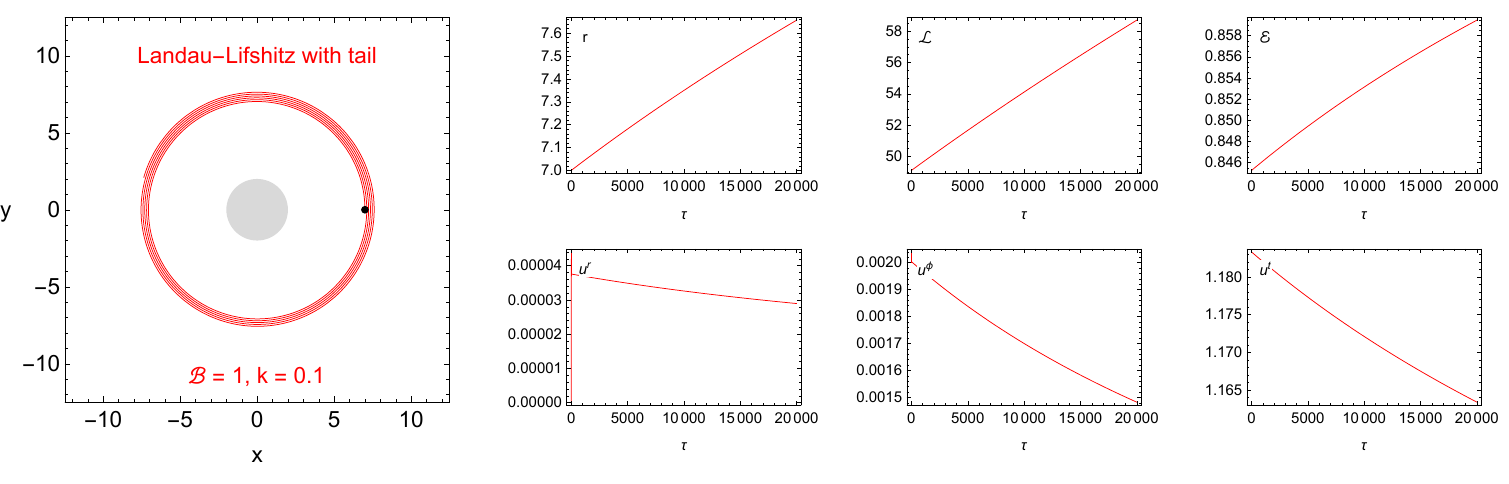}
\caption{The effect of the tail term. Charged particle trajectories for different $\BB$ and $k$, all starting at $r_{0}=7$. Top row: conservative Lorentz force; second row: pure tail effect; third and fourth rows: full radiation-reaction formalism including the tail term. Subfigures show the corresponding numerically calculated quantities.
}
\label{Fig10}
\end{figure*}

\subsection{DeWitt-Brehme vs. Landau-Lifshitz}

In this subsection we compare the DeWitt-Brehme (DB) and Landau-Lifshitz (LL) equations of motion for a charged particle in Schwarzschild spacetime with a uniform magnetic field. Since the tail term is identical for both equations, we first focus solely on the local, instantaneous radiation reaction, solving the tail term in a separate subsection below.  

We numerically solved the equations of motion for both DB (\ref{EOM1LD})--(\ref{EOM3LD}) and LL (\ref{EOM1LL})--(\ref{EOM3LL}) formulations. The resulting trajectories are presented in Fig.~\ref{Fig7}. 
To avoid runaway and pre-acceleration issues in the DB equations, we performed integrations backward in time, similarly to the approach taken in the flat spacetime case in \cite{Tur-Kol-Stu-Gal:2018:APJ:}. 
For direct comparison, we first integrated the LL equations forward in time, then used the final values from the LL integration as initial conditions for the backward integration of the DB equations. As a result, in Fig.~\ref{Fig7}, the starting point (indicated by a thick dot) for the LL trajectory corresponds to the endpoint of the DB trajectory, and vice versa. In all cases, the particle begins its motion in a circular orbit (i.e., without oscillations). We carried out this procedure for both positive and negative 
values of the magnetic parameter $\BB$. In all cases, the trajectories obtained from the DB and LL equations are in perfect agreement, demonstrating that both formulations yield identical results for the scenarios considered. 

As pointed out in~\cite{Tur-Kol-Stu-Gal:2018:APJ:} and \cite{Tur-Kol-Stu:2018:AN:}, particles with $\mathcal{B}<0$, that is, when the Lorentz force is directed toward the black hole, starting from a circular orbit will spiral into the black hole as a result of energy loss due to radiation reaction. By contrast, when $\mathcal{B}>0$, the situation is reversed. In this case, the guiding center (i.e., the center of the Larmor orbit) is effectively located at infinity. The radiation-reaction force then causes the particle to drift away from the black hole, toward this guiding center at infinity, leading to a gradual widening of the orbit. We will analyze this effect in more detail in the next section, including the role of the tail term. 

For now, we conclude that both the DB and LL equations of motion yield identical results when the tail term is neglected, both demonstrating the orbital-widening (OW) effect for $\BB>0$. Therefore, hereafter we proceed with the LL equations of motion for numerical integration, given their practical advantages for computation.

\subsection{Tail-term effect on particle orbits}

\subsubsection{Pure tail-term effect} 

Since the tail term is independent of external fields, one can isolate and investigate its pure effect by omiting both the Lorentz and local radiation-reaction forces. Specifically, we study the motion of a charged particle in Schwarzschild spacetime by numerically integrating the equations of motion with only the tail term included, as given by Eqs.~(\ref{tail_r})--(\ref{tail_Gal_phi}). In Fig.~\ref{Fig18} we present results of numerical integration showing the influence of the tail term on the motion of the charged particle for different values of $k$. The starting point of the particle is $r_{0}=7$. The figure shows that for a particle initially on a stable circular orbit, the inclusion of the tail term causes a delayed but eventual inspiral into the black hole. The inspiral time is strongly dependent on the strength of the tail term, with larger $k$ resulting in a more rapid orbital decay.

\subsubsection{Full radiation-reaction formalism with the tail} 

In Fig.~\ref{Fig10} we present a systematic comparison of charged particle trajectories for various choices of the magnetic parameter $\mathcal{B}$ and the radiation parameter $k$, with all initial conditions set at $r = 7$. The right panels in each row present the evolution of the numerically computed dynamical quantities, such as $r$, $\mathcal{L}$, $\mathcal{E}$, $u^t$, $u^{r}$ and $u^\varphi$. 

Each row Fig.~\ref{Fig10} corresponds to a different dynamical regime. The first row illustrates motion under the conservative Lorentz force, serving as a reference case with $\mathcal{B} \neq 0$, $k = 0$. As expected, the particle undergoes a stable circular orbit, with all conserved quantities remaining constant throughout the evolution.  

The second row demonstrates the effect of the pure tail term ($\mathcal{B} = 0$, $k \neq 0$). Here, the trajectory exhibits a clear inspiral,  accompanied by a monotonic decrease in $r$, $\mathcal{L}$, and $\mathcal{E}$, indicating the loss of angular momentum and energy from the system due to radiation reaction effects. The third and fourth rows show the results obtained by including both the Lorentz force, radiation-reaction force and the tail term within the full radiation-reaction formalism, for two different sets of parameters.

The last two rows of Fig.~\ref{Fig10} are particularly illustrative. The third row displays the inspiral, whereas the fourth row depicts orbital widening for a larger value of $\mathcal{B}$. It is evident that the inspiral is driven by the tail term: in its absence, the particle instead undergoes outward spiraling, as we have already demonstrated in the last two rows of Fig.~\ref{Fig7}. For larger $\mathcal{B}$, the relative contribution of the tail diminishes, and orbital widening reappears. We discuss the effect of orbital widening in a separate section below.

\section{Orbital widening effect} \label{sec:OWeffect}

%
\begin{figure*}[htbp]
\centering
\includegraphics[width=1\hsize]{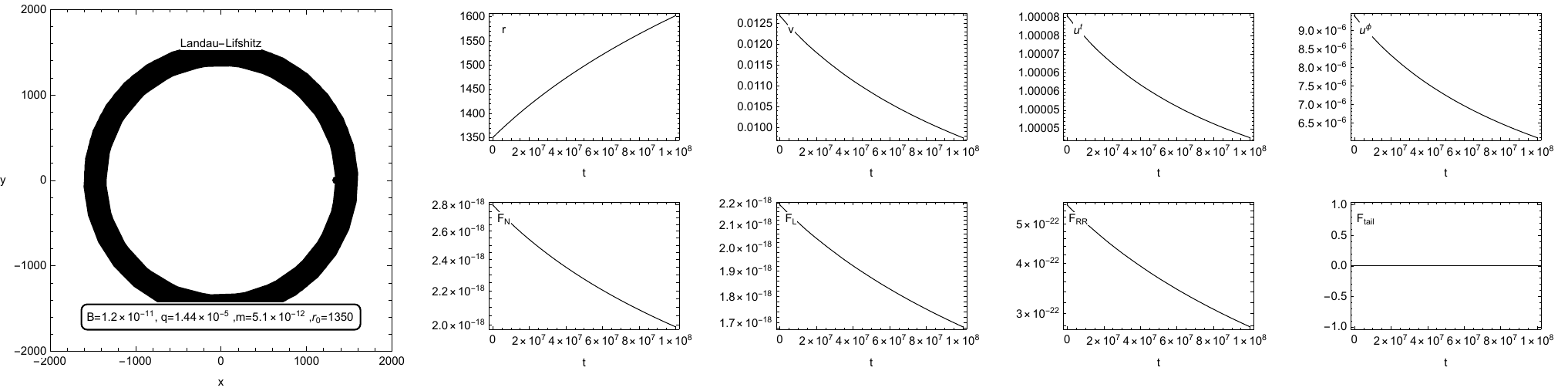}
\includegraphics[width=1\hsize]{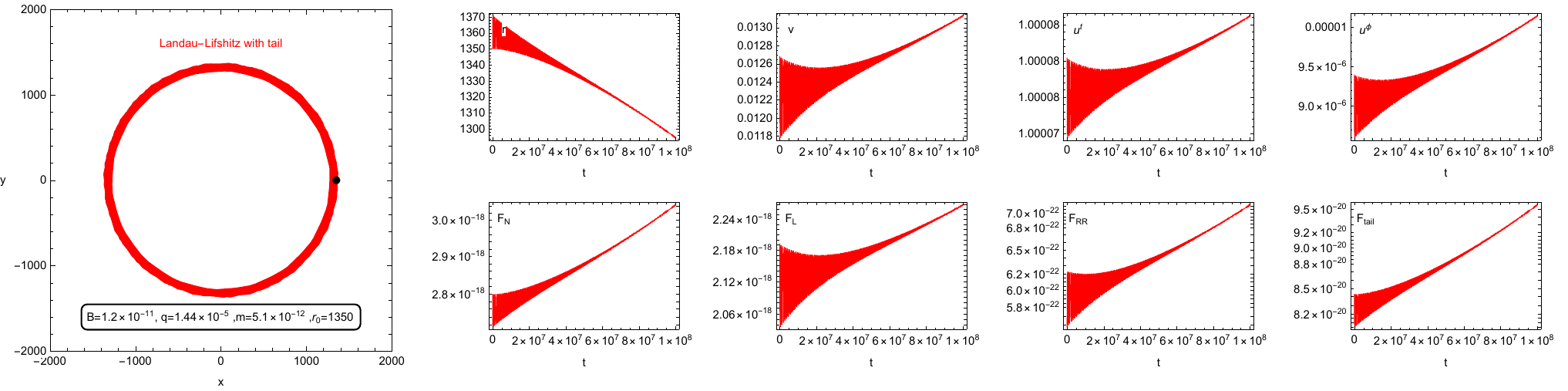}
\caption{
Radiating particle in a circular orbit within the Newtonian approximation, shown without (black) and with (red) the contribution of the tail term. The starting point of the particle is indicated by the black dot at $r_0 = 1350$. The parameter values $B$, $m$, and $q$ are taken from Table~I of ~\cite{San-Car-Nat:2023:PRD:}.
}
\label{Fig16}
\end{figure*}

\begin{figure*}[hbp]
\centering
\includegraphics[width=1\hsize]{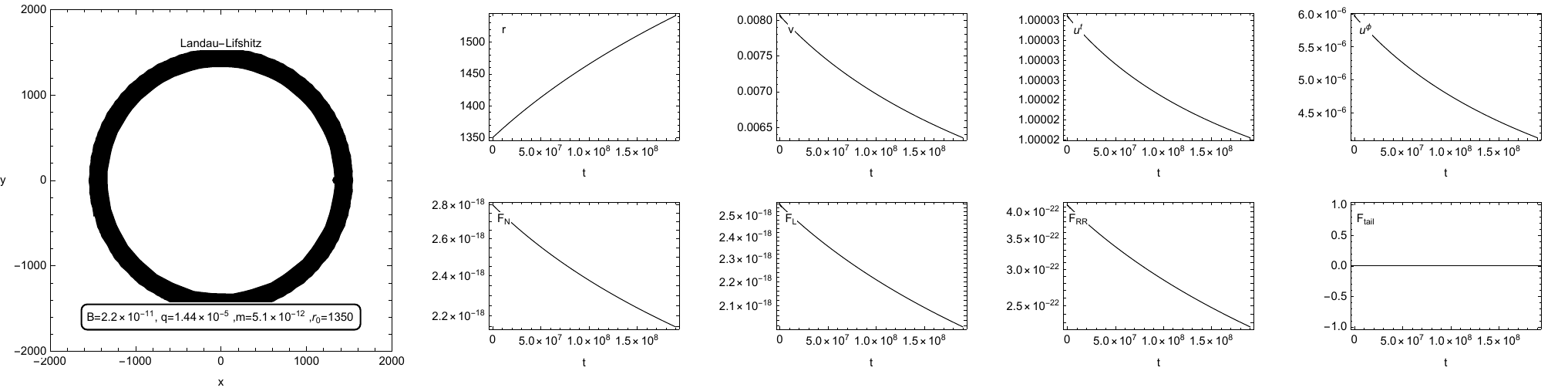}
\includegraphics[width=1\hsize]{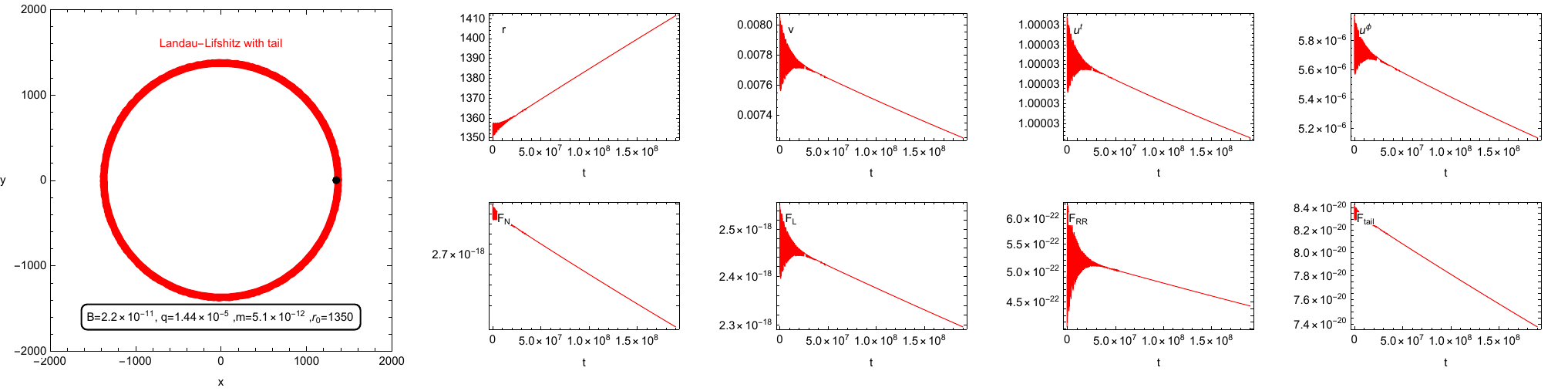}
\caption{ 
Same as in Fig.~\ref{Fig16} within the Newtonian approximation, but with an increased magnetic field strength of $B = 2.2 \times 10^{-11}$ instead of $B = 1.2 \times 10^{-11}$ in Fig.~\ref{Fig16}. The stronger field leads to occurence of the OW effect even in the presence of the tail term.}
\label{Fig17}
\end{figure*}

As pointed out in the previous section, for $\BB<0$ the particle spirals down to the black hole. But for $\BB>0$, we see that the orbit of the particle gets wider. This orbital widening effect (OW) has been observed for the first time in \cite{Tur-Kol-Stu-Gal:2018:APJ:,Tur-Kol-Stu:2018:AN:}. It was demonstrated that, in the presence of an external magnetic field, the radiation reaction force can lead to a systematic widening of the circular orbital radius, that is, the circular orbits shift outward from the black hole, provided that the Lorentz force acts outward (i.e., for  
$\BB L>0$, where $L$ is the azimuthal angular momentum of the particle). This effect occurs irrespective of the specific magnetic field configuration, being present for both homogeneous and dipole magnetic fields. It has been shown that while the kinetic energy of the particle decreases due to synchrotron radiation, the potential energy increases more rapidly as the orbit widens, resulting in a net increase in the energy and angular momentum measured at infinity.  

Recently, Santos et al. \cite{San-Car-Nat:2023:PRD:} critically examined the OW effect and argued that this phenomenon is an artifact arising from the neglect of the electromagnetic tail term in the equations of motion. Using a numerical example in the Newtonian limit, they claimed that the tail term must always be included whenever gravitational forces are relevant, and thus concluded that the OW effect should not occur. 

In this section, we study the OW effect in detail, provide counterexamples in Newtonian case and Schwarzschild spacetime, and demonstrate that the claim of Santos et al \cite{San-Car-Nat:2023:PRD:} is incorrect, showing that the OW effect is, in fact, a robust physical phenomenon.


%
\subsection{Orbital widening in the Newtonian limit}
\label{Newtonian_limit_OW}

We begin by considering the Newtonian approximation of the OW effect, which is valid in regimes with weak gravitational and magnetic fields and for particles moving at non-relativistic speeds. This approximation was studied in~\cite{San-Car-Nat:2023:PRD:}, where it was shown for a particular set of parameter values that the OW effect arises from neglecting the tail term in the equations of motion. Here, we show that the OW effect can persist even when the tail term is included, provided the magnetic field strength~$B$ is slightly increased, while keeping the other parameters identical to those in ~\cite{San-Car-Nat:2023:PRD:}. We refer to \cite{San-Car-Nat:2023:PRD:} for the equations of motion in Cartesian coordinates, setting the mass of the central object to $M=1$: 
\begin{align}
    m \frac{d u^{x}}{dt} = - \frac{m}{r^{3}}x + q B u^{y} - \frac{2 q^{3}}{3m^{2}} \left( q u^{x} B^{2} + \frac{m}{r^{3}}yB  \right) + \nonumber \\  + \frac{q^{2}}{ r^{3}} \left(\frac{x}{r} - \frac{2}{3} u^{x} \right), \\
    m \frac{d u^{y}}{dt} = - \frac{m}{r^{3}}y - q B u^{x} - \frac{2 q^{3}}{3m^{2}} \left( q u^{y} B^{2} - \frac{m}{r^{3}}xB  \right) + \nonumber \\  + \frac{q^{2}}{ r^{3}} \left(\frac{y}{r} - \frac{2}{3} u^{y} \right),
\end{align}
where $r=\sqrt{x^{2}+y^{2}}$ , $q$ - charge of the particle, $B$ - magnetic field intensity. 
In the equations above, we identify the first term on the right-hand side as the Newtonian force $F_{N}$; the second as the Lorentz force $F_{L}$; the third as the radiation reaction force $F_{RR}$; and the last term as the electromagnetic self-force $F_{\text{tail}}$.

We numerically solve the system of equations and demonstrate results in Figures~\ref{Fig16} and \ref{Fig17}. The parameters used in Fig.~\ref{Fig16} are taken directly from Table~I in~\cite{San-Car-Nat:2023:PRD:}. In Fig.~\ref{Fig17}, we consider the same setup but increase the magnetic field strength from $B = 1.2 \times 10^{-11}$ (used in Fig.~\ref{Fig16}) to $B = 2.2 \times 10^{-11}$. The figures illustrate the evolution of the particle’s radius $r$, speed $v$, the components $u^{t}$ and $u^{\phi}$ of its four-velocity, and the forces acting on the particle: the Newtonian force $F_{N}$, Lorentz force $F_{L}$, radiation reaction force $F_{RR}$, and the electromagnetic self-force $F_{\text{tail}}$, all as functions of time $t$.

The black trajectory represents the case without the tail term, while the red trajectory includes the tail term. As shown in Fig.~\ref{Fig16}, omitting the tail term causes the particle’s orbit to expand, whereas including it leads to a contraction of the orbital radius — consistent with the findings of~\cite{San-Car-Nat:2023:PRD:}. However, Fig.~\ref{Fig17} shows that a modest increase in the magnetic field strength~$B$ results in orbital widening even when the electromagnetic self-force is present. This enhancement of $B$ increases the initial magnitude of both the repulsive Lorentz force and the radiation reaction force, producing a net outward force that leads to the widening of the particle’s orbit. A dominance of the various components of forces over the tail term for radiating particles has also been discussed in Refs.~\cite{Tur-Kol-Stu-Gal:2018:APJ:, Tur-Kol-Stu:2018:AN:, Stuch-Kol-Tur:2024:Universe:,2020Univ....6...26S}. 

Our findings demonstrate that the OW effect persists even in the presence of the tail term, provided the magnetic field is sufficiently strong. This directly contradicts the conclusion of~\cite{San-Car-Nat:2023:PRD:}, which identified the absence of the tail term as the sole source of the OW effect. In the next section, we demonstrate that in realistic astrophysical settings, the magnetic field is typically strong enough that the tail term almost never eliminates the OW effect.

\subsection{Energy evolution in the Schwarzschild case}
%

\begin{figure}
\includegraphics[width=1\hsize]{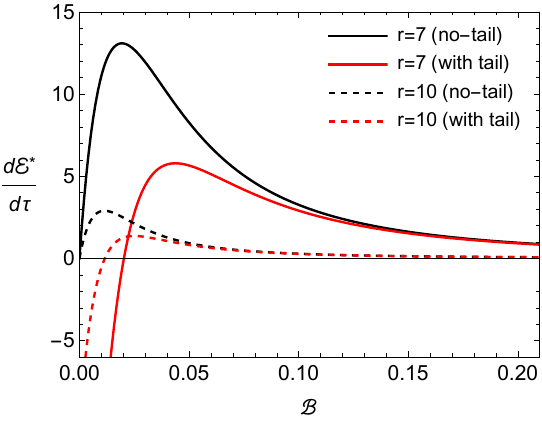}
\caption{Energy evolution of a radiating charged particle on a circular orbit, shown for different orbital radii ($r = 7$, $r = 10$) and in the presence (red) or absence (black) of the tail contribution. The quantity $\mathcal{E}^* = (10^5/k) \mathcal{E}$ denotes the rescaled energy. 
}
\label{fig-dE}
\end{figure}

Returning to the Schwarzschild case and using Eqs.~(\ref{Energy}) and (\ref{EOM1LL}), one can derive the energy loss for a radiating charged particle in a circular orbit 
\begin{equation}
    \frac{d \mathcal{E}}{d\tau} = - 4 k \mathcal{B}^2 \mathcal{E}^{3} + 2 k \mathcal{B} \, \mathcal{E} \left( 2 \mathcal{B} f + \frac{u^{\phi}}{r} \right)  + f \mathcal{F}^t_{tail}.
\label{Energy_loss}
\end{equation}
In the ultra-relativistic limit ($\mathcal{E} \gg 1$), this expression reduces to the well-known synchrotron loss formula, where only the first (negative) term on the right hand side dominates, leading to net energy loss. However, for $\mathcal{E} \lesssim 1$, which is typical for circular orbits, all three terms in Eq.~(\ref{Energy_loss}) can be relevant. 

It is evident that the second term remains positive for $\mathcal{B} > 0$. When this term becomes dominant compared to the other contributions, the OW effect manifests. 
For circular motion, the number of variables in Eq.~(\ref{Energy_loss}) can be reduced by using the expressions for the specific energy $\mathcal{E}_{\rm c.o.}$ and specific angular momentum $\mathcal{L}_{\rm c.o.}$ of a charged particle in a magnetized Schwarzschild background. These quantities at the equatorial plance are given by the following equations (see, e.g. \cite{Fro-Sho:2010:PHYSR4:,Kol-Stu-Tur:2015:CLAQG:}) 
\begin{eqnarray}
    \mathcal{E}_{\rm c.o.} &=&  \left( f(r) \left[ 1 + \left( \frac{\mathcal{L}_{\rm c.o.}}{r} - \mathcal{B} \, r \right)^2 \right] \right)^{1/2}, \\
    \mathcal{L}_{\rm c.o.} &=& 
\frac{-\mathcal{B} r^2 \pm r {\left( \mathcal{B}^2 r^2 (r - 2)^2 + r - 3 \right)^{1/2}}}{r - 3} .
\end{eqnarray}
Using the above expressions for $\mathcal{E}_{\rm c.o.}$ and $\mathcal{L}_{\rm c.o.}$, and substituting $u^{\phi}$ from Eq.~(\ref{Ang-mom}), the energy loss formula~(\ref{Energy_loss}) can be rewritten entirely in terms of the orbital radius $r$ and magnetic field parameter $\mathcal{B}$. 

In Fig.~\ref{fig-dE}, we plot the energy loss rate $d\mathcal{E}^*/d\tau=(10^5/k)d\mathcal{E}/d\tau$ as a function of $\mathcal{B}$ for different radii, and in both the presence and absence of the tail term. Several features can be observed from the figure:
\begin{itemize}[noitemsep]

    \item  In the absence of the tail term, the energy rate is always positive, indicating a net energy gain. This reflects the dominance of the second term in Eq.~(\ref{Energy_loss}) for ${\cal B}>0$ in the absence of the tail, which leads to the OW effect. 
    \item When the tail term is included and the magnetic field is weak ($\mathcal{B} \ll 1$), there exists a regime where the first and third (negative) terms dominate, leading to a net energy loss.
    \item As $\mathcal{B}$ increases, the energy rate becomes positive again, and the OW effect re-emerges. For both radii, the solutions with and without the tail term tend to converge at large $\mathcal{B}$, indicating a diminishing influence of the tail term in a weak gravitational and strong magnetic fields.
    \item This behavior is consistent with the origin of the tail term: it arises from spacetime curvature, while the magnetic field is uniform in the background spacetime considered here. Therefore, as the magnetic field strength increases, its effect outweighs the gravitational self-interaction.
    \item If the magnetic field strong enough for OW to occur, the energy gain is more pronounced at smaller radii, where the particle is closer to the black hole. 
\end{itemize}

These results demonstrate that the OW effect is a robust physical phenomenon, occurring in the case of repulsive Lorentz force and governed primarily by the interplay between magnetic field strength and orbital radius. The contribution of the tail term is most relevant in weak fields and diminishes at larger $\mathcal{B}$. The global conservation of energy for point particles undergoing radiation reaction was studied by Quinn and Wald \cite{Quinn-Wald:1999:PRD:}.

\begin{figure*}[htbp]
\centering
\fbox{%
        \begin{minipage}{0.98\textwidth}
            \centering
    \includegraphics[width=0.95\hsize]{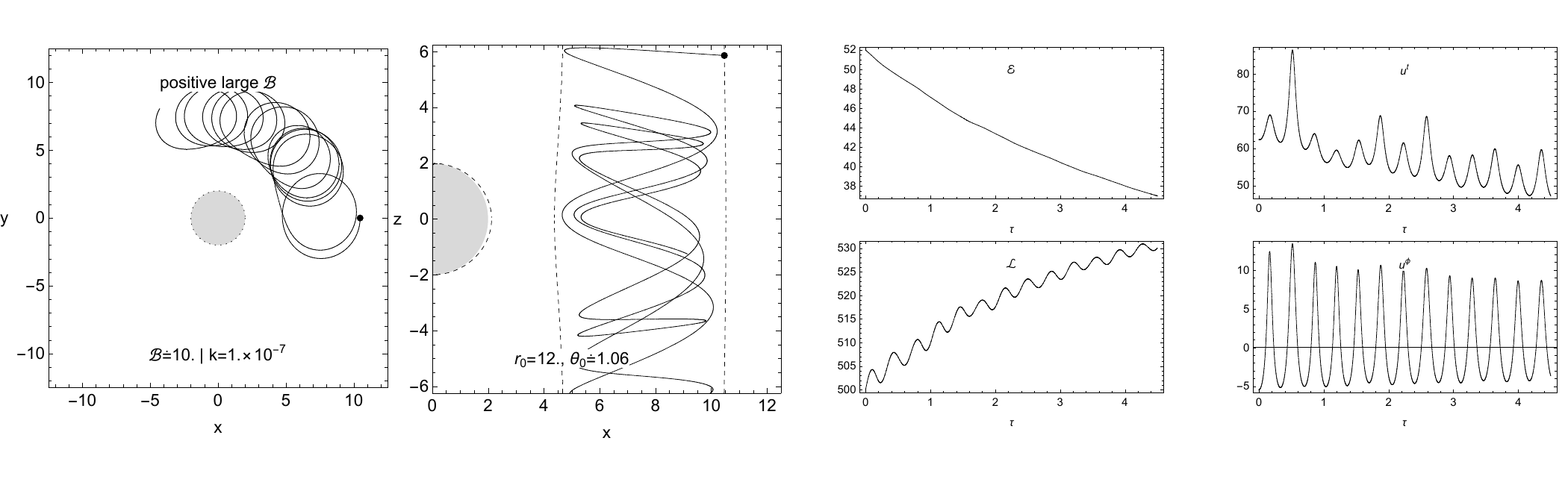}
    \includegraphics[width=0.95\hsize]{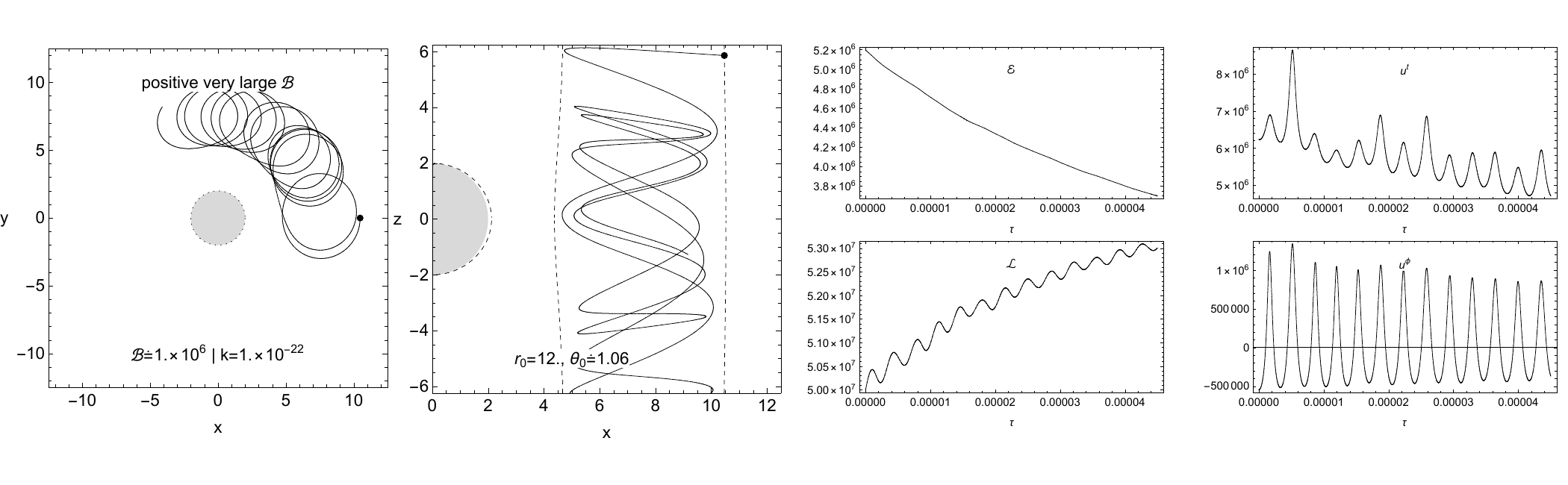}
        \end{minipage}%
    }
    \vspace{2mm}
    \fbox{%
        \begin{minipage}{0.98\textwidth}
            \centering
    \includegraphics[width=0.95\hsize]{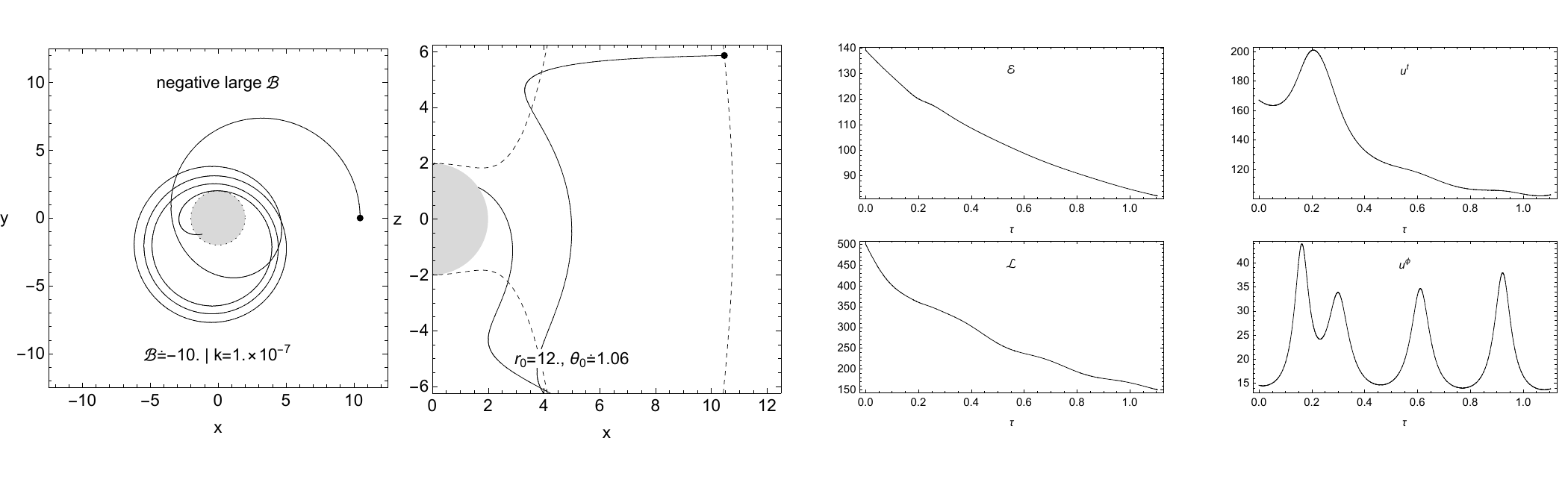}
    \includegraphics[width=0.95\hsize]{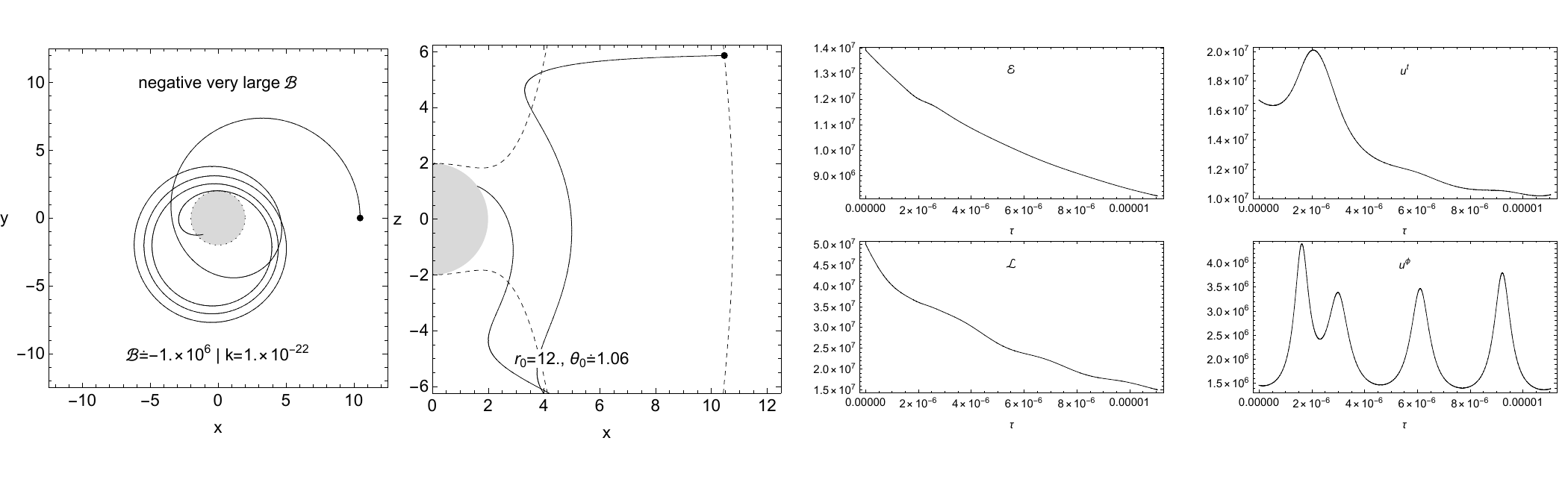}
        \end{minipage}%
    }
\caption{
Symmetry of trajectories of radiating charged particles for extremely different values of $\mathcal{B}$ and $k$, shown for both positive (top panels) and negative (bottom panels) $\mathcal{B}$ cases. In each panel the particle is released from the same initial radius $r_{0}=12$. Despite the large disparity in $\mathcal{B}$ and $k$, the orbital motion and corresponding dynamical quantities remain indistinguishable, illustrating the scaling symmetry discussed in Sec.~\ref{sec:astro}.} 
\label{Fig8}
\end{figure*}

\section{Astrophysical relevance}\label{sec:astro} 

\subsection{Scaling symmetry} 

In many studies of charged particle dynamics, both in this work and in the broader literature, representative values of the parameters are often chosen for illustration of qualitative effects. Such values, however, may not correspond directly to realistic astrophysical environments. In our case, the relevant dimensionless parameters are the magnetic field parameter $\mathcal{B}$ and the radiation-reaction parameter $k$. For the purposes of numerical simulations, we have considered moderate values such as $\mathcal{B}\sim 1$ and $k \lesssim 1$. In realistic astrophysical situations, however, the hierarchy is extreme: typically $\mathcal{B} \gg 1$, while $k \ll 1$. This naturally raises the question: can one be confident that the qualitative features we have identified at moderate parameter values persist in the physically relevant regime where the parameters differ by many orders of magnitude?

To address this, we uncover an interesting and nontrivial symmetry in the equations of motion. Remarkably, the trajectories of radiating charged particles become indistinguishable under a rescaling of parameters, provided the following relations are satisfied:
\begin{equation} \label{eq:scaling}
    \mathcal{L} = \mathcal{L}_{0} \mathcal{B}, 
    \quad 
    k = k_{0} \mathcal{B}^{-3}, 
    \quad 
    t = t_{0} \mathcal{B}^{-1}.
\end{equation}
Here $(\mathcal{L}_{0},k_{0},t_{0})$ denote the reference values, and $\mathcal{B}$ acts as the scaling parameter. In other words, the symmetry ensures that a trajectory computed for moderate values of $(\mathcal{B},k)$ can be mapped onto a physically realistic trajectory with $\mathcal{B}\gg 1$ and $k \ll 1$, by rescaling the orbital angular momentum, the strength of radiation reaction, and the temporal coordinate according to the relations above.

The scaling symmetry described above is illustrated in Fig.~\ref{Fig8}, where we compare trajectories of radiating charged particles for extremely different values of the magnetic parameter $\mathcal{B}$ and radiation-reaction parameter $k$. In each case the initial radius is fixed at $r_{0}=12$. The values of $(\mathcal{B},k)$ are chosen such that they differ by many orders of magnitude, yet satisfy the scaling relations of Eq.~(\ref{eq:scaling}). 

Remarkably, the orbital motion and the time evolution of the dynamical quantities (energy $\mathcal{E}$, angular momentum $\mathcal{L}$, and angular velocity $u^{\phi}$) are indistinguishable across the paired simulations. For example, the comparison between $\mathcal{B}=10$ with $k=10^{-7}$ and $\mathcal{B}=10^{5}$ with $k=10^{-22}$ shows that, after rescaling, both the spatial trajectories and the oscillatory structure of the kinematical variables coincide. This demonstrates that the symmetry is exact within numerical accuracy, even when the parameters differ by over fifteen orders of magnitude. 

The origin of the scaling symmetry, empirically discovered in this work, as well as its possible limitations, requires further detailed investigation, which lies beyond the scope of the present paper and is left for future work. 

The physical implication is important: simulations performed with moderate values of $\mathcal{B}$ and $k$, which are computationally tractable, faithfully represent the motion in realistic astrophysical conditions, where $\mathcal{B}$ is extremely large and $k$ is extremely small. Thus, the qualitative behavior reported in the previous sections, including the manifestation of the OW effect and the role of the tail term, can be extrapolated to astrophysical regimes.

\subsection{Magnitudes of forces}

\begin{table}[]
\begin{center}
\begin{tabular}{l  l  l l  l  l}
\hline
 & \quad $F_{\rm L}$ \quad & \quad $F_{\rm tail}$ \quad & \quad $F_{\rm RR1}$ \quad & \quad $F_{\rm RR2}$  \quad \\	
 B [G]  & \quad $\cb$ & \quad $k$ & \quad ${}k\cb$ & \quad ${}k\cb^2$ \\	
\hline
 $10^{15.8}$ & $\sim10^{19}$ & $\sim{}10^{-19}$ & $\sim1$ & $\sim10^{19}$ \\ 
 $10^{12}$ & $\sim10^{15}$ & $\sim{}10^{-19}$ &$\sim10^{-4}$ & $\sim10^{11}$\\  
 $10^8$ & $\sim10^{11}$ & $\sim{}10^{-19}$ &$\sim10^{-8}$ & $\sim 10^{3}$ \\
$10^{6.4}$ & $\sim10^{9}$ & $\sim{}10^{-19}$ &$\sim10^{-9}$ & $\sim1$ \\
 $10^4$ & $\sim10^{7}$ & $\sim{}10^{-19}$ &$\sim 10^{-12}$ & $\sim10^{-5}$ \\ 
 $10^0$ & $\sim10^{3}$ & $\sim{}10^{-19}$ &$\sim10^{-16}$ & $\sim 10^{-13}$ \\
 $10^{-2.9}$ & $\sim1$ & $\sim{}10^{-19}$ &$\sim 10^{-19}$ & $\sim10^{-19}$ \\
 $10^{-5}$ & $\sim10^{-2}$ & $\sim{}10^{-19}$ & $\sim 10^{-21}$ & $\sim10^{-23}$\\ 
\hline
\end{tabular}
\caption{Magnitudes of different forces acting on radiating charged particle in the dimensionless form of Eq. (\ref{LL}) for different values of magnetic field strengths. The estimates correspond to a relativistic electron in the vicinity of a stellar-mass BH with $M= 10~{M}_{\odot}$. Electron electromagnetic self-force (tail term) and gravitational interaction is the same for all cases $F_{\rm tail}\sim{}10^{-19}$ and $F_{\rm G}=1$. 
\label{TabForce}}
\end{center}
\end{table}

\begin{figure}
\includegraphics[width=1\hsize]{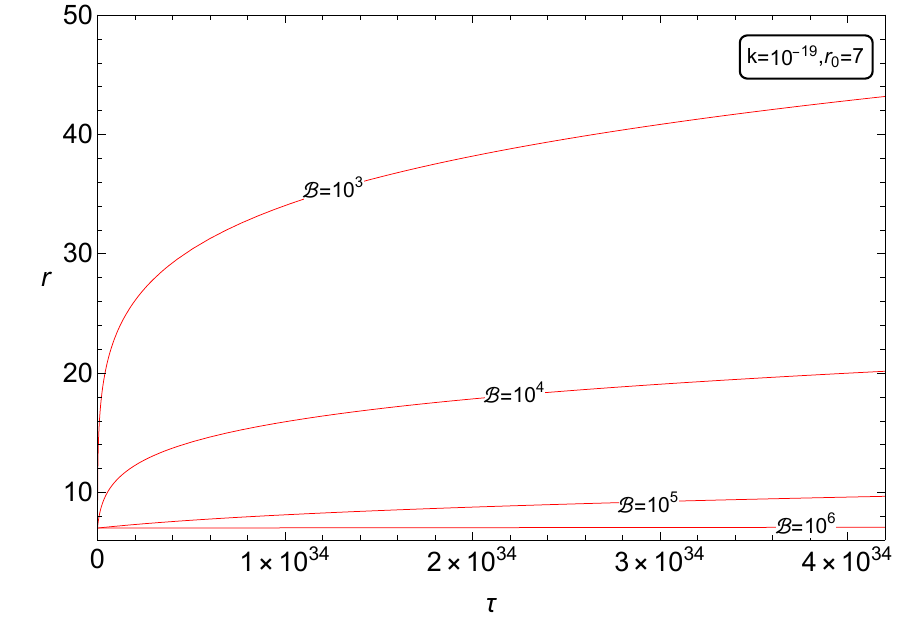}
\caption{ 
OW for realistic case.   
Radial coordinate evolution $r$ as a function of proper time $\tau$ for different values of $\mathcal{B}$, with fixed $k=10^{-19}$ corresponding to an electron. The particle starts from $r_{0}=7$, and the equations of motion are given by Eqs.~(\ref{EOM1LL}) - (\ref{EOM3LL}).
}
\label{Fig1}
\end{figure}

Table~\ref{TabForce} summarizes the relative magnitudes of the different forces entering the dimensionless form of Eq.~(\ref{LL}) for a relativistic electron near a stellar-mass black hole of mass $M=10\,M_{\odot}$. The estimates are given for a wide range of magnetic field strengths, from magnetar-level fields ($B\gtrsim 10^{14}\,{\rm G}$) down to interstellar values ($B\sim10^{-5}\,{\rm G}$).  

Several trends can be identified. First, the Lorentz force $F_{\rm L}\propto \mathcal{B}$ is by far the dominant term in the most of considered magnetic fields. 

Second, the two radiation-reaction contributions exhibit different scaling behaviors: $F_{\rm RR1}\sim k\mathcal{B}$ remains small across the entire range, while $F_{\rm RR2}\sim k\mathcal{B}^{2}$ can become comparable to unity at sufficiently large $\mathcal{B}$. For example, at $B\sim10^{15.8}\,{\rm G}$, $F_{\rm RR2}$ reaches $\sim10^{19}$, demonstrating that this term can rival the Lorentz force in extreme magnetar environments.  

The tail term $F_{\rm tail}\sim10^{-19}$ remains essentially constant and negligible compared to the Lorentz and radiation-reaction terms at astrophysical field strengths. This reflects its purely gravitational origin: while the electromagnetic contributions scale with $\mathcal{B}$, the tail term is set by spacetime curvature and is insensitive to the magnetic field. The gravitational force has been normalized to unity, $F_{\rm G}=1$, providing a natural reference scale in the dimensionless formulation.  

Taken together, the table emphasizes that in realistic astrophysical environments the dynamics of radiating charged particles is governed primarily by the Lorentz force and the quadratic radiation-reaction term, with the linear radiation-reaction contribution and the tail term playing a subdominant role. This hierarchy validates the use of scaling arguments introduced in the previous subsection and clarifies the physical conditions under which the OW effect can manifest near compact objects.

Fig.~\ref{Fig1} illustrates the radial evolution of a radiating electron for realistic values of the radiation–reaction parameter, fixed at $k=10^{-19}$, and for several values of the magnetic parameter $\mathcal{B}$. The particle starts from $r_{0}=7$. These results confirm that the OW effect persists for realistic parameter values, demonstrating its astrophysical relevance.

\section{Conclusions}
\label{Conclusions}

In this work, we have revisited the problem of electromagnetic radiation reaction for charged particles moving in the Schwarzschild spacetime immersed in a uniform magnetic field. Using both the full DeWitt–Brehme equations and their reduced-order covariant Landau–Lifshitz counterpart, we incorporated local and nonlocal self-force contributions of the electromagnetic tail term arising from spacetime curvature. 

We have confirmed that the higher-order DeWitt–Brehme and covariant Landau–Lifshitz formulations yield identical trajectories when the tail term is excluded, validating the use of the simpler LL equation for this system. By isolating the tail contribution and analyzing its conservative and dissipative components, we demonstrated agreement with previous analytic results and identified its limited dynamical role in astrophysical regimes. 

Most importantly, our results establish that the orbital widening effect is a genuine and robust physical phenomenon. 
Contrary to recent claims in the literature, we showed that OW persists even when the tail term is included, both in the Newtonian limit and in the full general relativistic treatment. By deriving a covariant energy evolution equation, we clarified the conditions under which OW occurs.

In the adopted formalism, radiating particles lose energy and momentum and drift toward their gyration center. In the case of a negative magnetic parameter, $\cb < 0$, the Lorentz force is attractive, acting toward the central object, placing the gyration center inside the black hole; in this regime, a radiating charged particle spirals inward toward the center. For a positive magnetic parameter, $\cb > 0$, the Lorentz force is repulsive with respect to the central object and counteracts black hole gravitational attraction, so the gyration center is located at infinity. The OW effect is possible only in the repulsive $\cb > 0$ case, leading to the increase of the total energy of the particle, despite continuous radiation losses, raising the question of the origin of the energy driving the OW process.

A static Schwarzschild spacetime offers, due to its symmetries, no mechanism for energy extraction: trajectories (photons) with strictly positive energies exist. Negative-energy photons, which may occur, for instance, in the ergosphere of a rotating Kerr black hole—where the radiative Penrose process can operate—are absent in the Schwarzschild case~\cite{Kol-Tur-Stu:2021:PRD:}. For an orbital-widening trajectory, we see the particle move toward its gyration center at infinity, and as it radiates photons with positive energy, its motion slows down. For such a trajectory, the orbital radius increases with time, and while its kinetic energy decreases, its total energy rises. Consequently, it moves upward in the effective potential toward the value corresponding to the rest-energy state at infinity, ${\cal E} = 1$. A charged particle initially on a bound orbit with ${\cal E}<1$ around the central object must acquire additional energy to reach infinity.

The proposed "strange" orbital widening trajectories are not unique to astrophysics and can be found in various bound systems, such as the Earth–Moon system, where the Moon gradually recedes from the Earth due to the transfer of Earth's internal rotational energy to the Moon via tidal forces. In the case of charged particles, the energy reservoir for orbital widening is likely the magnetic field generated by the particle’s circular motion; this generated magnetic field vanishes once the particle comes to rest. 
 
For a repulsive Lorentz force, $\cb > 0$, both the external uniform magnetic field and the field generated by the orbiting charged particle are aligned, and the particle’s magnetic field contributes positively to the total magnetic-field energy. Thus, an available energy reserve exists, which can be extracted as the charged particle is brought to rest and shifted to its gyration center at infinity. For an attractive Lorentz force, $\cb < 0$, the magnetic field generated by particle motion is oppositely oriented, corresponding to the classical case of a counterproductive interaction in which charged particles weaken the external magnetic field. In this regime, the motion of the charged particles reduces the total magnetic field energy, and no scope for orbital widening exists.

Interestingly, we find that in the presence of the tail term, the energy evolution equation can change sign for certain values of the magnetic parameter $\mathcal{B}$, given a fixed orbital radius. This implies the existence of nontrivial circular orbits which, despite emitting radiation and orbiting a black hole, maintain a constant orbital radius. 

We further uncovered a nontrivial scaling symmetry in the equations of motion, allowing us to map results across many orders of magnitude in parameter space. This symmetry enables us to extrapolate numerical findings obtained in idealized settings to realistic astrophysical scenarios involving electrons near realistic black holes.  It shows that the qualitative numerical examples demonstrated in the paper, such as the occurrence of the OW effect, the role of the tail term, and the overall structure of energy evolution, remains valid in the realistic astrophysical regimes near compact objects.

\section{Acknowledgments}
This work is supported by the Silesian University in Opava Grant No. SGS/24/2024. B.J. acknowledges the Moravian-Silesian Region Foundation for its support as part of the project “Support for Talented Doctoral Students at the Silesian University in Opava 2022”. AT and MK acknowledge support from the Czech Science Foundation Grant No.~\mbox{23-07043S}. 

\bibliography{reference}

@ARTICLE{2024JHEAp..44..500S,
       author = {{Stuchl{\'\i}k}, Zden{\v{e}}k and {Vrba}, Jaroslav and {Kolo{\v{s}}}, Martin and {Tursunov}, Arman},
        title = "{Radiative back-reaction on charged particle motion in the dipole magnetosphere of neutron stars}",
      journal = {Journal of High Energy Astrophysics},
         year = 2024,
        month = nov,
       volume = {44},
        pages = {500-530},
          doi = {10.1016/j.jheap.2024.11.006},
archivePrefix = {arXiv},
       eprint = {2412.04996},
 primaryClass = {astro-ph.HE},
       adsurl = {https://ui.adsabs.harvard.edu/abs/2024JHEAp..44..500S}
}

@article{Stuch-Kol-Tur:2024:Universe:,
    author = "Stuchl\'\i{}k, Zden\v{e}k and Kolo\v{s}, Martin and Tursunov, Arman and Gal\textquoteright{}tsov, Dmitri",
    title = "{On the Role of the Tail Term in Electromagnetic Radiation Reaction}",
    doi = "10.3390/universe10060249",
    journal = "Universe",
    volume = "10",
    number = "6",
    pages = "249",
    year = "2024"
}

@PHDTHESIS{Haas:2008:PHDThesis:,
       author = {{Haas}, Roland},
        title = "{Self-force on point particles in orbit around a Schwarzschild black hole}",
       school = {University of Guelph, Canada},
         year = 2008,
        month = jan,
       adsurl = {https://ui.adsabs.harvard.edu/abs/2008PhDT........26H}
}

@article{Wald:1978:PRD:,
  title = {Construction of Solutions of Gravitational, Electromagnetic, or Other Perturbation Equations from Solutions of Decoupled Equations},
  author = {Wald, Robert M.},
  journal = {Phys. Rev. Lett.},
  volume = {41},
  issue = {4},
  pages = {203--206},
  numpages = {0},
  year = {1978},
  month = {Jul},
  publisher = {American Physical Society},
  doi = {10.1103/PhysRevLett.41.203},
  url = {https://link.aps.org/doi/10.1103/PhysRevLett.41.203}
}

@article{Det-Whi:2004:LRR:,
   title={Self-force via a Green’s function decomposition},
   volume={67},
   ISSN={1089-4918},
   url={http://dx.doi.org/10.1103/PhysRevD.67.024025},
   DOI={10.1103/physrevd.67.024025},
   number={2},
   journal={Physical Review D},
   publisher={American Physical Society (APS)},
   author={Detweiler, Steven and Whiting, Bernard F.},
   year={2003},
   month=jan }

@ARTICLE{Jur-Stuch-Tur-Kol:2024:JCAP:,
       author = {{Juraev}, Bakhtinur and {Stuchl{\'\i}k}, Zden{\v{e}}k and {Tursunov}, Arman and {Kolo{\v{s}}}, Martin},
        title = "{Radiating particles accelerated by a weakly charged Schwarzschild black hole}",
      journal = {\jcap},
         year = 2024,
        month = sep,
       volume = {2024},
       number = {9},
          eid = {035},
        pages = {035},
          doi = {10.1088/1475-7516/2024/09/035},
archivePrefix = {arXiv},
       eprint = {2402.13797},
 primaryClass = {gr-qc},
       adsurl = {https://ui.adsabs.harvard.edu/abs/2024JCAP...09..035J}
}

@ARTICLE{Gal:1982:JPMG:,
       author = {{Gal'tsov}, D.~V.},
        title = "{Radiation reaction in the Kerr gravitational field}",
      journal = {Journal of Physics A Mathematical General},
         year = 1982,
        month = dec,
       volume = {15},
       number = {12},
        pages = {3737-3749},
          doi = {10.1088/0305-4470/15/12/025},
       adsurl = {https://ui.adsabs.harvard.edu/abs/1982JPhA...15.3737G}
}

@article{Quinn-Wald:1997:PRD:,
  title = {Axiomatic approach to electromagnetic and gravitational radiation reaction of particles in curved spacetime},
  author = {Quinn, Theodore C. and Wald, Robert M.},
  journal = {Phys. Rev. D},
  volume = {56},
  issue = {6},
  pages = {3381--3394},
  numpages = {0},
  year = {1997},
  month = {Sep},
  publisher = {American Physical Society},
  doi = {10.1103/PhysRevD.56.3381},
  url = {https://link.aps.org/doi/10.1103/PhysRevD.56.3381}
}

@article{War-Bara:2010:PRD:,
  title = {Self-force on a scalar charge in Kerr spacetime: Circular equatorial orbits},
  author = {Warburton, Niels and Barack, Leor},
  journal = {Phys. Rev. D},
  volume = {81},
  issue = {8},
  pages = {084039},
  numpages = {17},
  year = {2010},
  month = {Apr},
  publisher = {American Physical Society},
  doi = {10.1103/PhysRevD.81.084039},
  url = {https://link.aps.org/doi/10.1103/PhysRevD.81.084039}
}

@article{San-Car-Nat:2023:PRD:,
  title = {Electromagnetic radiation reaction and energy extraction from black holes: The tail term cannot be ignored},
  author = {Santos, Jo\~ao S. and Cardoso, Vitor and Nat\'ario, Jos\'e},
  journal = {Phys. Rev. D},
  volume = {107},
  issue = {6},
  pages = {064046},
  numpages = {8},
  year = {2023},
  month = {Mar},
  publisher = {American Physical Society},
  doi = {10.1103/PhysRevD.107.064046},
  url = {https://link.aps.org/doi/10.1103/PhysRevD.107.064046}
}

@article{DeWitt-DeWitt:1964:PPF:,
  title = {Falling charges},
  author = {DeWitt, C\'ecile Morette and DeWitt, Bryce S.},
  journal = {Physics Physique Fizika},
  volume = {1},
  issue = {1},
  pages = {3--20},
  numpages = {18},
  year = {1964},
  month = {Jul},
  publisher = {American Physical Society},
  doi = {10.1103/PhysicsPhysiqueFizika.1.3},
  url = {https://link.aps.org/doi/10.1103/PhysicsPhysiqueFizika.1.3}
}

@article{Quinn-Wald:1999:PRD:,
  title = {Energy conservation for point particles undergoing radiation reaction},
  author = {Quinn, Theodore C. and Wald, Robert M.},
  journal = {Phys. Rev. D},
  volume = {60},
  issue = {6},
  pages = {064009},
  numpages = {20},
  year = {1999},
  month = {Aug},
  publisher = {American Physical Society},
  doi = {10.1103/PhysRevD.60.064009},
  url = {https://link.aps.org/doi/10.1103/PhysRevD.60.064009}
}

@ARTICLE{Hob:1968:AP:,
       author = {{Hobbs}, J.~M.},
        title = "{A vierbein formalism of radiation damping}",
      journal = {Annals of Physics},
         year = 1968,
        month = mar,
       volume = {47},
       number = {1},
        pages = {141-165},
          doi = {10.1016/0003-4916(68)90231-5},
       adsurl = {https://ui.adsabs.harvard.edu/abs/1968AnPhy..47..141H}
}

@ARTICLE{Tur-Kol-Stu-Gal:2018:APJ:,
       author = {{Tursunov}, Arman and {Kolo{\v{s}}}, Martin and {Stuchl{\'\i}k}, Zden{\v{e}}k and {Gal'tsov}, Dmitri V.},
        title = "{Radiation Reaction of Charged Particles Orbiting a Magnetized Schwarzschild Black Hole}",
      journal = {\apj},
         year = 2018,
        month = jul,
       volume = {861},
       number = {1},
          eid = {2},
        pages = {2},
          doi = {10.3847/1538-4357/aac7c5},
archivePrefix = {arXiv},
       eprint = {1803.09682},
 primaryClass = {gr-qc},
       adsurl = {https://ui.adsabs.harvard.edu/abs/2018ApJ...861....2T}
}

@ARTICLE{Kol-Tur-Stu:2021:PRD:,
       author = {{Kolo{\v{s}}}, Martin and {Tursunov}, Arman and {Stuchl{\'\i}k}, Zden{\v{e}}k},
        title = "{Radiative Penrose process: Energy gain by a single radiating charged particle in the ergosphere of rotating black hole}",
      journal = {\prd},
         year = 2021,
        month = jan,
       volume = {103},
       number = {2},
          eid = {024021},
        pages = {024021},
          doi = {10.1103/PhysRevD.103.024021},
archivePrefix = {arXiv},
       eprint = {2010.09481},
 primaryClass = {gr-qc},
       adsurl = {https://ui.adsabs.harvard.edu/abs/2021PhRvD.103b4021K}
}

@ARTICLE{Poi-Pou-Veg:2011:LRR:,
       author = {{Poisson}, Eric and {Pound}, Adam and {Vega}, Ian},
        title = "{The Motion of Point Particles in Curved Spacetime}",
      journal = {Living Reviews in Relativity},
         year = 2011,
        month = sep,
       volume = {14},
       number = {1},
          eid = {7},
        pages = {7},
          doi = {10.12942/lrr-2011-7},
archivePrefix = {arXiv},
       eprint = {1102.0529},
 primaryClass = {gr-qc},
       adsurl = {https://ui.adsabs.harvard.edu/abs/2011LRR....14....7P}
}

@ARTICLE{Smi-Wil:1980:PRD:,
       author = {{Smith}, A.~G. and {Will}, Clifford M.},
        title = "{Force on a static charge outside a Schwarzschild black hole}",
      journal = {\prd},
         year = 1980,
        month = sep,
       volume = {22},
       number = {6},
        pages = {1276-1284},
          doi = {10.1103/PhysRevD.22.1276},
       adsurl = {https://ui.adsabs.harvard.edu/abs/1980PhRvD..22.1276S}
}

@ARTICLE{DeW-Bre:1960:AP:,
       author = {{DeWitt}, Bryce S. and {Brehme}, Robert W.},
        title = "{Radiation damping in a gravitational field}",
      journal = {Annals of Physics},
         year = 1960,
        month = feb,
       volume = {9},
       number = {2},
        pages = {220-259},
          doi = {10.1016/0003-4916(60)90030-0},
       adsurl = {https://ui.adsabs.harvard.edu/abs/1960AnPhy...9..220D}
}

@ARTICLE{Tur-Kol-Stu:2018:AN:,
       author = {{Tursunov}, A.~A. and {Kolo{\v{s}}}, M. and {Stuchl{\'\i}k}, Z.},
        title = "{Orbital widening due to radiation reaction around a magnetized black hole}",
      journal = {Astronomische Nachrichten},
         year = 2018,
        month = jun,
       volume = {339},
       number = {5},
        pages = {341-346},
          doi = {10.1002/asna.201813502},
archivePrefix = {arXiv},
       eprint = {1806.06754},
 primaryClass = {gr-qc},
       adsurl = {https://ui.adsabs.harvard.edu/abs/2018AN....339..341T}
}

@ARTICLE{Spohn:2000:EPL:,
       author = {{Spohn}, H.},
        title = "{The critical manifold of the Lorentz-Dirac equation}",
      journal = {EPL (Europhysics Letters)},
         year = 2000,
        month = may,
       volume = {50},
       number = {3},
        pages = {287-292},
          doi = {10.1209/epl/i2000-00268-x},
archivePrefix = {arXiv},
       eprint = {physics/9911027},
 primaryClass = {physics.acc-ph},
       adsurl = {https://ui.adsabs.harvard.edu/abs/2000EL.....50..287S}
}

@article{Stu-Kol:2016:EPJ:,
	author = {{Stuchl\'{\i}k, Zdenek} and {Kolos, Martin}},
	title = {Acceleration of the charged particles due to chaotic scattering in the combined black hole gravitational field and asymptotically uniform magnetic field},
	DOI= "10.1140/epjc/s10052-015-3862-2",
	url= "https://doi.org/10.1140/epjc/s10052-015-3862-2",
	journal = {Eur. Phys. J. C},
	year = 2016,
	volume = 76,
	number = 1,
	pages = "32",
	month = "",
}

@String { and      = { and } }

@book{Landau:1975:CTP2:,
    author = "Landau, L. D. and Lifschits, E. M.",
    title = "{The Classical Theory of Fields}",
    isbn = "978-0-08-018176-9",
    publisher = "Pergamon Press",
    address = "Oxford",
    series = "Course of Theoretical Physics",
    volume = "Volume 2",
    year = "1975"
}

@ARTICLE{2020Univ....6...26S,
       author = {{Stuchl{\'\i}k}, Zden{\v{e}}k and {Kolo{\v{s}}}, Martin and
         {Kov{\'a}{\v{r}}}, Ji{\v{r}}{\'\i} and {Slan{\'y}}, Petr and
         {Tursunov}, Arman},
        title = "{Influence of Cosmic Repulsion and Magnetic Fields on Accretion Disks Rotating around Kerr Black Holes}",
      journal = {Universe},
         year = 2020,
        month = jan,
       volume = {6},
       number = {2},
        pages = {26},
          doi = {10.3390/universe6020026},
       adsurl = {https://ui.adsabs.harvard.edu/abs/2020Univ....6...26S}
}

@ARTICLE{Wald:1974:PHYSR4:,
   author = {{Wald}, R.~M.},
    title = "{Black hole in a uniform magnetic field}",
  journal = {\prd},
     year = 1974,
    month = sep,
   volume = 10,
    pages = {1680-1685},
      doi = {10.1103/PhysRevD.10.1680}
}

@ARTICLE{Poisson:2004:LRR:,
       author = {{Poisson}, Eric},
        title = "{The Motion of Point Particles in Curved Spacetime}",
      journal = {Living Reviews in Relativity},
         year = 2004,
        month = may,
       volume = {7},
       number = {1},
          eid = {6},
        pages = {6},
          doi = {10.12942/lrr-2004-6},
archivePrefix = {arXiv},
       eprint = {gr-qc/0306052},
 primaryClass = {gr-qc},
       adsurl = {https://ui.adsabs.harvard.edu/abs/2004LRR.....7....6P}
}

@ARTICLE{Kol-Stu-Tur:2015:CLAQG:,
       author = {{Kolo{\v{s}}}, Martin and {Stuchl{\'\i}k}, Zden{\v{e}}k and {Tursunov}, Arman},
        title = "{Quasi-harmonic oscillatory motion of charged particles around a Schwarzschild black hole immersed in a uniform magnetic field}",
      journal = {Classical and Quantum Gravity},
         year = 2015,
        month = aug,
       volume = {32},
       number = {16},
          eid = {165009},
        pages = {165009},
          doi = {10.1088/0264-9381/32/16/165009},
archivePrefix = {arXiv},
       eprint = {1506.06799},
 primaryClass = {gr-qc},
       adsurl = {https://ui.adsabs.harvard.edu/abs/2015CQGra..32p5009K}
}

@ARTICLE{Fro-Sho:2010:PHYSR4:,
   author = {{Frolov}, V.~P. and {Shoom}, A.~A.},
    title = "{Motion of charged particles near a weakly magnetized Schwarzschild black hole}",
  journal = {\prd},
archivePrefix = "arXiv",
   eprint = {1008.2985},
 primaryClass = "gr-qc",
     year = 2010,
    month = oct,
   volume = 82,
   number = 8,
      eid = {084034},
    pages = {084034},
      doi = {10.1103/PhysRevD.82.084034}
}

@article{Bay-Hus:1976:PRD:,
  title = {Nonuniqueness of physical solutions to the Lorentz-Dirac equation},
  author = {Baylis, W. E. and Huschilt, J.},
  journal = {Phys. Rev. D},
  volume = {13},
  issue = {12},
  pages = {3237--3239},
  numpages = {0},
  year = {1976},
  month = {Jun},
  publisher = {American Physical Society},
  doi = {10.1103/PhysRevD.13.3237},
  url = {https://link.aps.org/doi/10.1103/PhysRevD.13.3237}
}

@ARTICLE{Tur-Stu-Kol:2016:PRD:,
       author = {{Tursunov}, Arman and {Stuchl{\'\i}k}, Zden{\v{e}}k and {Kolo{\v{s}}}, Martin},
        title = "{Circular orbits and related quasiharmonic oscillatory motion of charged particles around weakly magnetized rotating black holes}",
      journal = {\prd},
         year = 2016,
        month = apr,
       volume = {93},
       number = {8},
          eid = {084012},
        pages = {084012},
          doi = {10.1103/PhysRevD.93.084012},
archivePrefix = {arXiv},
       eprint = {1603.07264},
 primaryClass = {gr-qc},
       adsurl = {https://ui.adsabs.harvard.edu/abs/2016PhRvD..93h4012T}
}

\end{document}